\documentclass[a4paper,fleqn,usenatbib]{mnras}

\usepackage{newtxtext,newtxmath}

\usepackage[T1]{fontenc}
\usepackage{ae,aecompl}

\usepackage{graphicx}	
\usepackage{amsmath}	
\usepackage{amssymb}	
\usepackage{xspace}

\usepackage{etoolbox}

\usepackage{color}

\newcommand{\Siggas}{\ensuremath{\Sigma_\mathrm{g}}\xspace}
\newcommand{\rhos}{\ensuremath{\rho_\text{s}}\xspace}
\newcommand{\uf}{\ensuremath{u_\text{f}}\xspace}

\newcommand{\Sigdust}{\ensuremath{\Sigma_\mathrm{d}}\xspace}

\title[Dusty proto-planetary discs radius evolution]{The time evolution of dusty protoplanetary disc radii: observed and physical radii differ}

\author[G. P. Rosotti et al.]{Giovanni P. Rosotti,$^{1,2}$\thanks{E-mail: rosotti@ast.cam.ac.uk}, Marco Tazzari$^{1}$, Richard A. Booth$^{1}$, Leonardo Testi$^{3}$, \newauthor Giuseppe Lodato$^{4}$ and Cathie Clarke$^{1}$
\\
$^{1}$Institute of Astronomy, Madingley Road, Cambridge, CB3 0HA, UK\\
$^{2}$Leiden Observatory, Leiden University, P.O.~Box 9531, NL-2300~RA Leiden, the Netherlands\\
$^{3}$European Southern Observatory, Karl-Schwarzschild-Str 2, D-85748 Garching, Germany\\
$^{4}$Universit\`a degli Studi di Milano, Via Giovanni Celoria 16, I-20133 Milano, Italy
}

\date{Accepted 2019 April 17. Received 2019 April 12; in original form 2018 November 5}

\pubyear{2019}

\begin{document}
\label{firstpage}
\pagerange{\pageref{firstpage}--\pageref{lastpage}}
\maketitle

\begin{abstract}
Proto-planetary disc surveys conducted with ALMA are measuring disc radii in multiple star forming regions. The disc radius is a fundamental quantity to diagnose whether discs undergo viscous spreading, discriminating between viscosity or angular momentum removal by winds as drivers of disc evolution. Observationally, however, the sub-mm continuum emission is dominated by the dust, which also drifts inwards, complicating the picture. In this paper we investigate, using theoretical models of dust grain growth and radial drift, how the radii of dusty viscous proto-planetary discs evolve with time. Despite the existence of a sharp outer edge in the dust distribution, we find that the radius enclosing most of the dust \textit{mass} increases with time, closely following the evolution of the gas radius. This behaviour arises because, although dust initially grows and drifts rapidly onto the star, the residual dust retained on Myr timescales is relatively well coupled to the gas. Observing the expansion of the dust disc requires using definitions based on high fractions of the disc \textit{flux} (e.g. 95 per cent) and very long integrations with ALMA, because the dust grains in the outer part of the disc are small and have a low sub-mm opacity. We show that existing surveys lack the sensitivity to detect viscous spreading. The disc radii they measure do not trace the mass radius or the sharp outer edge in the dust distribution, but the outer limit of where the grains have significant sub-mm opacity. We predict that these observed radii should shrink with time.
\end{abstract}




\section{Introduction}

Planet formation takes place in proto-planetary discs, which provide the building blocks (gas and solids) to assemble the numerous planetary systems observed around main sequence stars (see e.g. \citealt{WinnFabrycky2015} for a review). The way the disc evolves affects the availability of the building blocks of planet formation and is therefore of fundamental importance to understanding planet formation.

Thanks to the transformational capabilities of the Atacama Large Millimetre Array (ALMA), it is now becoming possible to observe large samples of discs of different ages, gathering essential statistics to understand how disc evolution takes place. The two quantities that are most readily accessible are the sub-mm continuum disc fluxes (normally considered to be a proxy for the mass under the optically thin assumption) and radii. Several ALMA surveys have already been published, reporting measurements of masses \citep{BarenfeldFirst,Pascucci2016,AnsdellLupusI,AnsdellSigmaOri,2018MNRAS.478.3674R} and radii \citep{BarenfeldSizes,AnsdellLupusII,Cox2017,Cieza2018} in different star forming regions. As a counterpart, these surveys have already sparked \citep{Rosotti2017,Lodato2017,Mulders2017} a renewed theoretical interest in understanding the mechanisms regulating disc evolution.



One way in which these surveys could shed light on our understanding of disc evolution is by testing the theories that aim to explain the observational evidence \citep[e.g.,][]{1988ApJ...330..350B,1995ApJ...452..736H} that discs accrete. It has been hypothesised \citep{LyndenBellPringle74} that proto-planetary discs evolve under the influence of an effective \textit{viscosity}, for convenience often parametrised with the convention of \citet{ShakuraSunyaev1973} and generally thought to be caused by the magneto-rotational instability (MRI) (e.g., \citealt{BalbusHawley91}; see \citealt{ArmitageReview,TurnerReview} for recent reviews). An alternative, emerging picture \citep{SuzukiInutsuka2009,Fromang2013,BaiStone2013} is one in which \textit{disc winds} drive accretion by carrying away angular momentum rather than transporting it through the disc. 

A fundamental prediction of viscous theory is that the angular momentum of the disc should be conserved. Therefore, while the bulk of the mass moves inwards and is eventually accreted onto the star, some parts of the disc must move outwards to conserve angular momentum. This leads to \textit{viscous spreading}: discs get larger with time. In principle, this could be tested observationally by comparing the disc radii in regions of different age, and in this way one could assess whether discs evolve viscously or under the influence of winds. 

Intriguingly, \citet{Tazzari2017} recently reported that the discs in Lupus are larger and less luminous than the discs in Taurus, a younger region, in line with the expectations of viscous spreading. This result is still a matter of debate since \citet{Tripathi} and \citet{Andrews2018}, using results from the Submilliter Array (SMA) in Taurus and ALMA in Lupus, do not find any statistically significant discrepancy between the two regions.

There is a big caveat when straightforwardly interpreting disc radii inferred from the sub-mm continuum emission as a probe of viscous spreading. This experiment should be performed using an optically thin gas emission line (such as C$^{18}$O, rather than optically thick like $^{12}$CO) capable of tracing how the gas mass in the disc is distributed. Even with the sensitivity improvements of ALMA, however, this remains a challenge due to the long observing time requested. At the time of writing, there is no significant sample of measured disc radii in C$^{18}$O and observational studies are still relying on the dust component of discs. This is much easier to access in the sub-mm since it dominates the opacity and the emission is considered to be optically thin, allowing one to trace the spatial distribution of the solid component of the disc. Many theoretical works however \citep{Weidenschilling1977,TakeuchiLin2002,Birnstiel2014} have highlighted that the dynamics of the dust is different from the dynamics of the gas due to the so-called \textit{radial drift}. As a result of the drag force from the gas, the dust loses angular momentum, spiralling inwards towards the star. While quantifying the importance of radial drift for observations is difficult \citep{Hughes2008,Facchini2017} due to opacity and excitation effects, there is now putative observational evidence \citep{Isella2012,deGregorioMonsalvo2013,Andrews2016,Cleeves2016} of this phenomenon, since in many discs the dust emission is more compact than the gas emission as predicted by theoretical models \citep{Birnstiel2014}. To complicate the picture even further, radial drift is a process that depends sensitively on the grain size; therefore, its observational consequences are deeply interwoven with the processes controlling grain growth \citep{2007ApJ...671.2091G,Birnstiel2009}.


Given the importance of radial drift, it is perhaps surprising that the evolution of the disc dust radius in a viscously evolving disc has never been the subject of a comprehensive theoretical study. The purpose of this paper is to address this gap and to study whether the evolution of the dust disc radius is set by viscous spreading (and can therefore be used as a probe of viscous evolution) or by the dust processes (namely growth and radial drift). Note that, in contrast to previous investigations \citep{Birnstiel2014,Facchini2017}, the focus of this study is not on the mismatch in disc radii between gas and dust \textit{at a given time}, but on how the dust radii should \textit{evolve in time}.

The magnitude of radial drift is a sensitive function of the grain size and it is thus important to consider grain growth to address this problem. To this effect, we employ current state of the art models of grain growth \citep{Birnstiel2012}, a significant difference from previous studies like \citet{TakeuchiClarkeLin2005} who did not evolve the grain size with time. We then compute synthetic sub-mm surface brightness profiles from the models and investigate their radii as observed by ALMA.

The paper is structured as follows: in section \ref{sec:methods} we discuss the methods and assumptions in our modelling and in section \ref{sec:example} we illustrate a particular case in detail. In the following two sections we present our results when we vary the parameters of the problem, respectively in section \ref{sec:mass_radius_all} for the mass evolution and in section \ref{sec:flux} for the flux evolution. Finally in section \ref{sec:discussion} we discuss the observational implications for current and future disc surveys and we draw our conclusions in section \ref{sec:conclusions}.

\section{Methods}
\label{sec:methods}

In this paper we evolve the dust and gas in the disc on secular timescales. We use the viscous evolution equations for the gas, while for the dust we use the simplified treatment of grain growth described in \citet{Birnstiel2012}. This treatment has the advantage of being computationally cheap to evaluate, yet it reproduces correctly the results of significantly more computationally expensive models of grain growth \citep{2008A&A...480..859B,2010A&A...513A..79B} that solve the coagulation equation at each point in the disc. As a post-processing step, we compute the opacity at ALMA wavelengths resulting from the dust properties obtained from the grain growth model and use it to generate synthetic surface brightness profiles. These profiles can then be compared with real observations.

\subsection{Disc evolution}

The code we use has been presented in \citet{Booth2017}; we refer the reader to that paper for a detailed description and here we only summarise the most important aspects. Following \citet{Birnstiel2012}, at each radius we evolve two dust populations: a population of small grains, with a grain size of 0.1 $\mu$m, and one of large grains that comprises most of the mass. We set the mass fraction in each of the two populations using the coefficients quoted in \citet{Birnstiel2012}, which are calibrated to reproduce the results of detailed grain growth models. We set the maximum grain size\footnote{{In the rest of the manuscript we will often refer simply to "grain size" rather than "maximum grain size" for simplicity.}} following the relations of \citet{Birnstiel2012} to take into account the effects of grain growth. In brief, the grain size at each radius is set either by fragmentation, or by radial drift, whichever is the lowest. In the former case the grain size is given by 

\begin{equation}
a_\mathrm{frag} = f_\mathrm{f}\, \frac{2}{3 \, \pi} \, \frac{\Siggas}{\rhos \alpha} \, \frac{\uf^2}{c_s^2},
\label{eq:a_frag}
\end{equation}

where $\Siggas$ is the local gas surface density, $\rhos$ is the grain bulk density, $c_s$ is the gas sound speed, $f_\mathrm{f}$ is an order of unity dimensionless factor (calibrated against more detailed simulation; we fix it to 0.37 following \citealt{Birnstiel2012}) and $a$ denotes the radius of a dust grain. The two most important parameters in setting the grain size are $\alpha$, the \citet{ShakuraSunyaev1973} parametrization of the viscosity (see later) and $\uf$, the fragmentation velocity, which in this paper we set to 10 m/s. Since the relative velocity of collisions between dust grains due to turbulence increases with size, the fragmentation limit corresponds to the maximum size that allows grains to collide without fragmenting.  In the opposite regime, the maximum grain size is given by

\begin{equation}
a_\mathrm{drift} = f_\mathrm{d} \, \frac{2 \, \Sigdust}{\pi \, \rhos}\frac{V_\mathrm{k}^2}{c_s^2} \, \gamma^{-1},
\label{eq:a_drift}
\end{equation}
where $f_\mathrm{d}$ is another order of unity factor (which we set to 0.55 following \citealt{Birnstiel2012}), $\Sigdust$ is the surface density of the dust, $V_\mathrm{k}$ is the Keplerian velocity and $\gamma$ is the absolute value of the local power-law slope of the gas pressure $P(r,t)= c_s^2 \ensuremath{\rho_\mathrm{g}}\xspace(r,t)$ (more formally, $| \mathrm{d}\log P / \mathrm{d} \log r |$), where $\ensuremath{\rho_\mathrm{g}}\xspace=\Siggas/\sqrt{2 \pi} H$ is the gas density in the midplane. The drift limit corresponds to the limit in which the dust grains radially drift as fast as they grow.

Regarding the time evolution of grain size, we notice that most of the quantities in Equations \ref{eq:a_frag} and \ref{eq:a_drift} do not evolve with time. Therefore, at each given radius the grain size in the fragmentation dominated case depends only on the surface density of the gas, while in the second depends only on the dust surface density.

Once the grain size has been calculated, we use the one-fluid approach described in \citet{LaibePrice2014} to compute the dust radial drift velocity. This approach allows us to consider both the drag force of the gas on the dust and the feedback of the dust onto the gas, which could potentially be a significant effect \citep{2018MNRAS.479.4187D}. In practice, because of fast radial drift the dust-to-gas ratio decreases so quickly that the feedback is not significant. The fundamental parameter controlling the dynamics \citep[e.g.,][]{Weidenschilling1977} is the Stokes number $St$:
\begin{equation}
St = \frac{\pi}{2} \frac{a \rho_s}{\Siggas},
\end{equation}
which is proportional to the grain size $a$ and inversely proportional to the gas surface density $\Siggas$. Grains with $St \sim 1$ drift the fastest, grains with $St \ll 1$ are well coupled to the gas and grains with $St \gg 1$  do not move radially.

In contrast to \citet{Booth2017}, in this paper we are not concerned with the inner disc, but rather we focus on the outer disc. For this reason we do not include viscous heating, which is a significant effect only in the inner $\sim$ 1 au. We rather opt to simply prescribe the temperature as a radial power-law. We used a two layer model \citep{1997ApJ...490..368C} to calibrate the temperature for a solar mass star to $40 \ (r/10 \mathrm{au})^{-0.5} \ \mathrm{K}$, corresponding to an aspect ratio $H/r = 0.033$ at 1 au.

In terms of the viscosity, we assume that the viscous torque only acts on the gas. We use the \citet{ShakuraSunyaev1973} parametrization to set the magnitude of the viscosity coefficient $\nu = \alpha c_s H$ at each radius, where $\alpha$ is the \citet{ShakuraSunyaev1973} dimensionless parameter, $c_s$ is the sound speed (which we compute from the prescribed temperature assuming a mean molecular weight of 2.4) and $H=c_s/\Omega$ is the disc scale-height. With our choice of the temperature, the viscosity $\nu \propto r$. 

In this paper we explore the dependence of viscous spreading on the value of $\alpha$. In particular, we consider the values $\alpha=10^{-2}, 10^{-3}$ and $10^{-4}$, which encompass the typical range of variation of viscosity given at the upper end by the MRI and at the lower end by hydrodynamical instabilities. In addition, we also consider a higher value of $\alpha=0.025$ for illustrative purposes; while it is not clear whether the MRI is able to drive such an efficient angular momentum transport, especially at large radii, it is certainly worth exploring how the predictions would change in this case. We shall see how a relatively modest variation of a factor 2.5 in viscosity can make a significant difference to the predictions. To give a reference value, with our choice of the temperature profile the viscous time $t_\nu= r^2/3 \nu$ is 0.5 Myr at 10 au if $\alpha=10^{-3}$. With the values of $\alpha$ we employ, most of the disc is in the fragmentation dominated regime for $\alpha \geq 10^{-2}$ (though with $\alpha=10^{-2}$ the disc switches to the drift limited regime after $\sim$1 Myr of evolution, see section \ref{sec:sensitivity_viscosity}).

As for the initial conditions, we use the analytical solution of \citet{LyndenBellPringle74} corresponding to the chosen viscosity law: 
\begin{equation}
\Sigma \propto r^{-1} \exp(-r/r_1),
\label{eq:sigma}
\end{equation}
where $r_1$ is a scaling radius (containing $1-1/e \sim 63$ per cent of the mass of the disc). In what follows we experiment with different values of $r_1$, using the values 10, 30 and 80 au. We set the normalization of the surface density depending on the initial disc mass $M_d = 2 \pi \int \Sigma r \mathrm{d} r$, which we set to $0.1 \ M_\odot$. The initial mass has little impact in terms of the radius evolution because both in the fragmentation and drift dominated regimes the Stokes number is independent of the surface density. In the interest of simplicity, we will therefore use a single value for all the models presented in this paper. Finally, we assume a uniform dust-to-gas ratio of $10^{-2}$ throughout the disc in the initial conditions.

\subsection{Surface brightness calculation}

\begin{figure}
\includegraphics[width=\columnwidth]{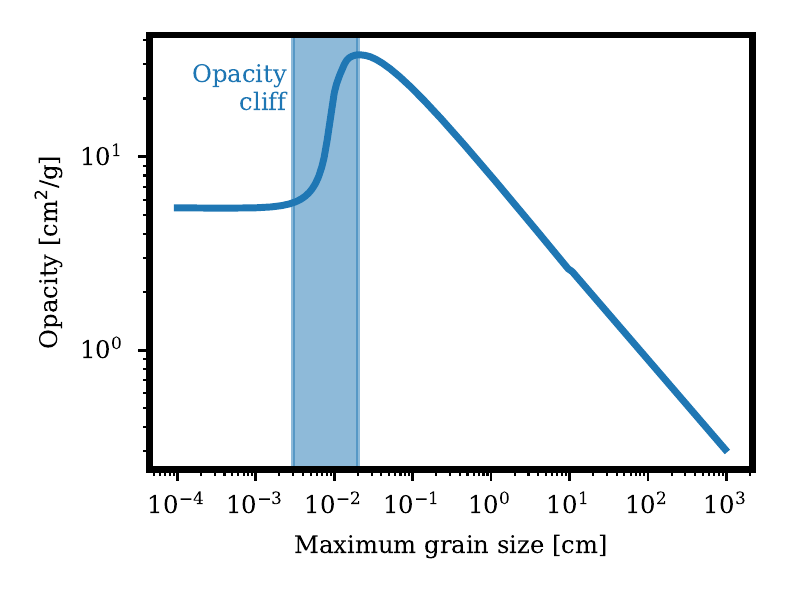}
\caption{The dust opacity at 850 $\mu$m we employ in this paper as a function of the maximum grain size, assuming that at each radius the number density of dust grains is a power-law with exponent -3.5. We marked on the figure the location of the ``opacity cliff'' (where the opacity steeply drops by one order of magnitude over a small range of variation in grain size; see text).}
\label{fig:opacity}
\end{figure}

As a post-processing step, we compute the sub-mm surface brightness of the disc as

\begin{equation}
S_b (R) = B_\nu (T(R) ) [1-\exp(-\kappa_\nu \Sigma_\mathrm{dust})],
\end{equation}
where $B_\nu$ is the Planck function, $\kappa_\nu$ the dust opacity and $\Sigma_\mathrm{dust}$ the surface density of the dust component. For simplicity we assume face-on discs. Given that the emission in the sub-mm is coming from {a thin layer in the disc mid-plane}, we do not expect inclination to introduce any significant difference in what we discuss in this paper. For comparison with the ALMA surveys, we compute the surface brightness in band 7, i.e. at 850 $\mu$m. We compute opacity as in \citet{2016A&A...588A..53T} following models by \citet{Natta:2004yu} and \citet{Natta:2007ye}, using the Mie theory for compact spherical grains with a simplified version of the volume fractional abundances in \citet{1994ApJ...421..615P}, assuming a composition of 10\% silicates, 30\% refractory organics, and 60\% water ice. As discussed in the previous section, the model of grain growth computes the maximum grain size $a_\mathrm{max}$ at each radius. To turn this into an opacity, we assume that at each radius the grain size distribution is a power-law $n(a)\propto a^{-q}$ for $a_\mathrm{min}\leq a\leq a_\mathrm{max} $ with an exponent $q=3.5$ \citep{1977ApJ...217..425M}. We show the resulting opacity as a function of the maximum grain size in \autoref{fig:opacity}. It is worth reflecting on the shape of this curve, in particular on the abrupt change in opacity that happens around the characteristic size of {0.02 cm} where the maximum opacity is attained. Moving towards smaller grains, the opacity decreases steeply (a factor of $\sim$ 10) over a narrow range of grain sizes. The opacity decreases also for larger grains, but the decrease is significantly shallower on this side. We shall see that the net result is effectively to make parts of the disc ``invisible'' as the grain size drops below the critical value. We will refer to the sharp drop in opacity as the ``\textit{opacity cliff}''. Quantitatively, the exact shape of the opacity cliff (the critical dust size and the opacity drop) depend slightly on the exact dust composition; for simplicity, in this paper we consider only one composition. On the other hand, the opacity cliff disappears completely if one considers ``fluffy'' rather than compact grains \citep{Kataoka:2014aa}. The growth model we use in this paper by construction considers compact grains and therefore we do not consider this possibility further.

\subsection{Radius determination}
\label{sec:radius_det}

Since the disc is a continuous structure, assigning it a characteristic scale is somehow arbitrary. The problem is mitigated for the initial conditions, where the simple functional form of the surface density allows us to define the scaling radius previously mentioned. However, as the disc evolves, the surface density takes a different functional form and this no longer applies.

For this reason, we opt to use a simple definition that we can apply irrespectively of the precise functional form of the surface density: for any given tracer (gas or dust) we define the disc radius as the radius that encloses a fixed fraction of the total disc mass at any given time. There is still a degree of arbitrariness in deciding which fraction to use. For consistency with the definition of scaling radius (see \autoref{eq:sigma}), we will use the 63 per cent fraction, though we note that the results are relatively insensitive to the precise value. 

Observations however do not measure the disc surface density, but its surface brightness. For this reason we define also an \textit{observed} disc radius using the synthetic surface brightness profile. In analogy with the mass radius, we define it as the radius enclosing a given fraction of the total disc flux. While earlier observational papers employed physical models of the surface density to fit the observations, it has been recently realised that this is a degenerate problem since the grain size is a function of radius. For this reason, two recent surveys have used a similar criterion based on a given fraction of the flux: \citet{Tripathi} and \cite{Andrews2018} used the 68 per cent flux radius ({note this is different from the 63 per cent we use for the mass}) and \citet{AnsdellLupusII} 90 per cent. 
In what follows we will experiment with different fractions of the total flux; as we shall see, in contrast to the mass radius, the behaviour depends on the adopted fraction.

For brevity, we will call in the rest of the paper ``\textit{mass radius}'' the radius definition based on the disc surface density and ``\textit{flux radius}'' the radius definition based on the disc surface brightness.

\section{A worked example}
\label{sec:example}

To better illustrate our results, we first present a worked example in detail. Subsequently, we show how the results change when varying the parameters of the disc.

\subsection{General features}

\label{sec:example_general}

\begin{figure}
\includegraphics[width=\columnwidth]{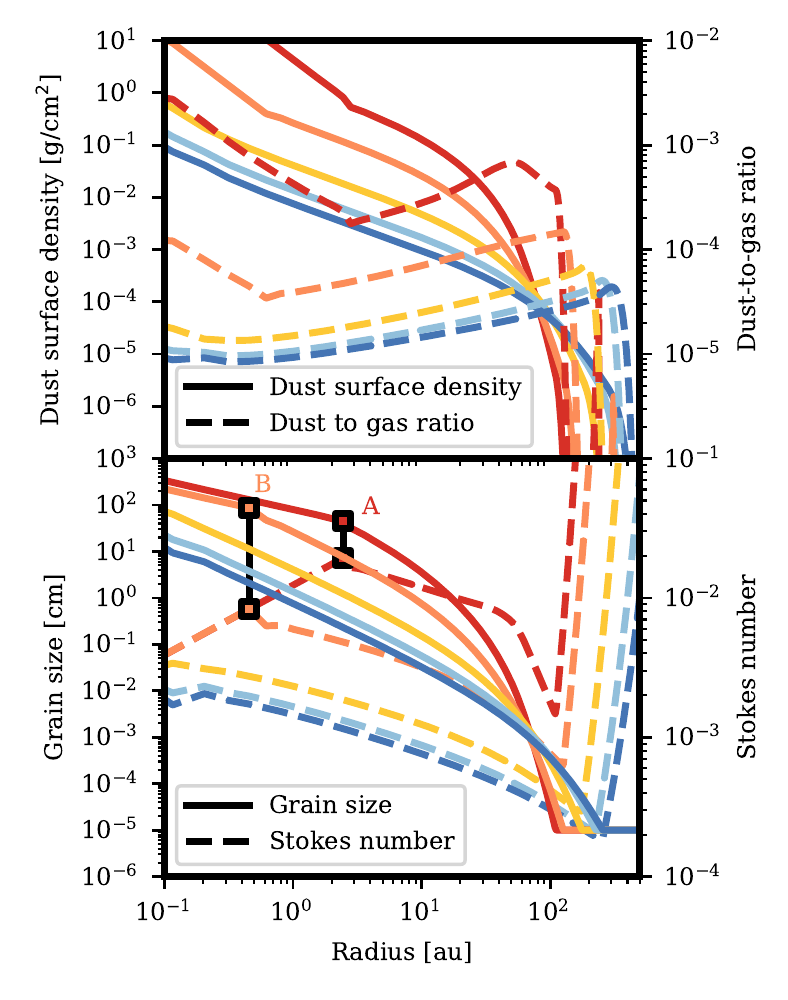}
\caption{\textbf{Top panel}: evolution in time (0.1, 0.3, 1, 2 and 3 Myr, with colors ranging from red to blue) of the dust surface density (solid lines). To better highlight the sharp dust outer edge, we plot also the dust-to-gas ratio (dashed lines). \textbf{Bottom panel}: evolution of the maximum grain size (solid lines) and of the Stokes number (dashed line). We have highlighted the transition radius between fragmentation and drift dominated regime for the first two timesteps with the squares and the letters A and B.}
\label{fig:summary_0.001_0.1_10}
\end{figure}

We choose as fiducial model a disc with $\alpha=10^{-3}$ and an initial radius of 10 au. In this model the value of viscosity is intermediate inside the admissible range from MRI and well below the existing upper limits from direct measures of the turbulence \citep{2018ApJ...856..117F}; with the chosen initial radius the initial accretion rate is $\sim 10^{-7} \ M_\odot \ \mathrm{yr}^{-1}$, in line with the highest measured accretion rates of class II objects. The initial viscous timescale of the disc is 0.5 Myr, consistent with the analysis of \citet{Lodato2017} in the Lupus star forming region. The parameters of this  model are also very similar to those of \citet{Owen2010}, which, when coupled with X-ray photo-evaporation, reproduce the observed disc lifetimes and mass accretion rate distribution.

\autoref{fig:summary_0.001_0.1_10} shows in the top panel the dust surface density and dust-to-gas ratio at different times (0.1, 0.3, 1, 2 and 3 Myr, with colors ranging from red to blue), and in the bottom panel the grain size (solid lines) and corresponding Stokes numbers (dashed lines). Similar results have already been presented in \citet{Birnstiel2012} but we choose to summarise them here in order to facilitate the understanding of the radius evolution. We stress in particular the following features: 

\begin{itemize}
\item The dust depletes on a very fast timescale; by the end of the simulation the dust-to-gas ratio has a typical value of $10^{-5}$. This is the well known fact that, because of radial drift, discs experience a large dust depletion.

\item The grain size follows two distinct behaviours depending on the radius; the transition radius between the two regimes can be recognised as a knee in the grain size or Stokes number, as we have highlighted in the bottom panel of \autoref{fig:summary_0.001_0.1_10} with the squares and the letters A and B for the first two timesteps. While $a_\mathrm{frag} (r) \propto \Siggas/c_s^2$, $a_\mathrm{drift} (r) \propto \Sigdust V_\mathrm{k}^2/c_s^2$ and has therefore a steeper dependence with radius even if the surface densities of gas and dust have the same slope. As a consequence in the inner part of the disc the grains are in the fragmentation dominated regime. On the contrary, in the outer part of the disc the relevant regime is the drift dominated one.

\item As time passes, both the fragmentation and drift dominated grain sizes (Equations \ref{eq:a_frag} and \ref{eq:a_drift}) become smaller as a result of gas and dust accretion onto the star: therefore the dust grain size at each radius is a decreasing function of time (for what concerns the Stokes number, note that in the fragmentation dominated regime the Stokes number is fixed with time). Because the dust is preferentially depleted with respect to the gas, with time the drift dominated regime encompasses a larger part of the disc, even at small radii. In fact, in this model the transition radius between the two regimes moves to {a distance smaller than 1 au} already after 0.2 Myr.

\begin{figure}
\includegraphics[width=\columnwidth]{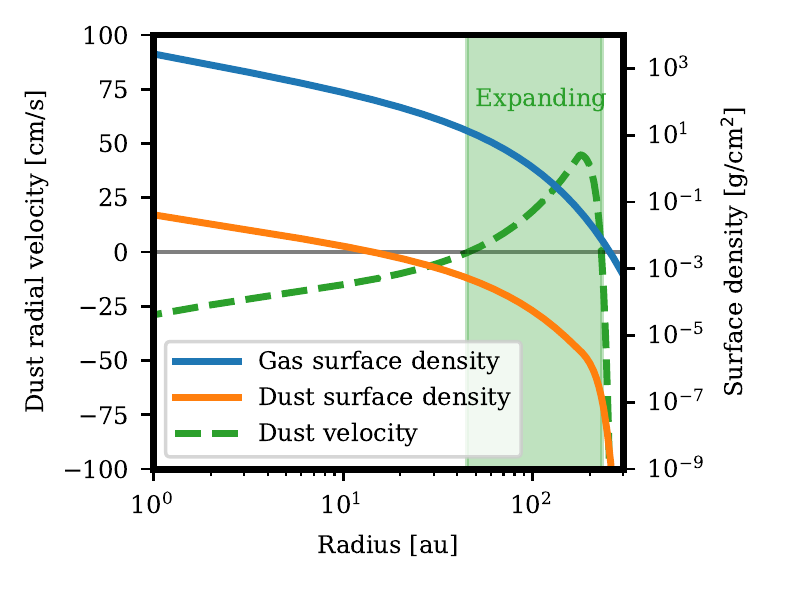}
\caption{The gas and dust surface densities at $t=$1 Myr and the dust radial velocity. The sharp dust outer edge seen in \autoref{fig:summary_0.001_0.1_10} is a result of the fast drift velocity in the outer part of the disc. While the dust is drifting inwards also close to the star, there is an intermediate region (shaded on the plot) where the grains are relatively well coupled to the gas and move outwards. This regions is driving the expansion of the dust disc.}
\label{fig:dust_sigma_vel}
\end{figure}

\item \textit{At any given time} the surface density of the dust presents a sharp outer edge: see the sharp drop in dust-to-gas ratio at large radii. Note instead how the dust-to-gas ratio inside the disc is almost flat. We will not try to define quantitatively what the outer edge is, but to illustrate why this feature develops, we plot in \autoref{fig:dust_sigma_vel} the dust drift velocity at $t=1$ Myr. For reference we include also on the same plot the gas and dust surface densities. The sharp outer edge is sculpted by the strong \textit{inwards} velocity in the outer part of the disc, a consequence of the gas surface density becoming very steep in the outer parts (due to the exponential dependence with radius). This feature was the focus of the investigation of \citet{Birnstiel2014}.

\autoref{fig:dust_sigma_vel} also shows that at intermediate radii, before the sharp outer edge, there is a region of the disc where the velocity is directed \textit{outwards}, a consequence of the fact that the Stokes number in this part of the disc is small enough that the radial drift velocity is (in absolute magnitude) smaller than the gas velocity. We shall see in the next section \ref{sec:evol_time_mass} how it is this part of the disc that drives the evolution of the disc radius \textit{with time}.

\end{itemize}

\subsection{Time evolution of the mass radius}
\label{sec:evol_time_mass}

\begin{figure}
\includegraphics[width=\columnwidth]{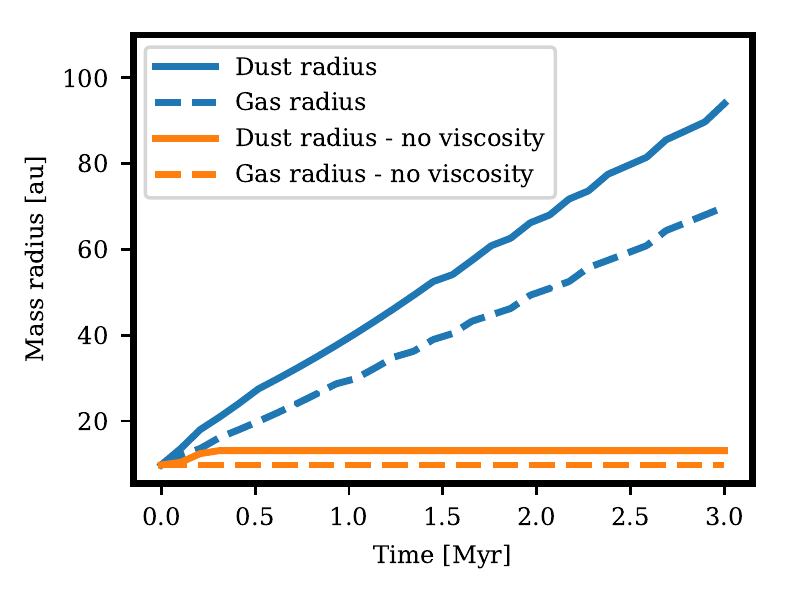}
\caption{Evolution of the dust and gas mass radius for the fiducial case. We plot also the results of a control run in which we do not take into account viscous evolution, confirming that the expansion of the dust disc is due to viscosity.}
\label{fig:rdisc_example_evol}
\end{figure}

Having summarised the general features of the dust evolution, we can now move to the objective of this paper, investigating how the mass radius evolves with time. \autoref{fig:rdisc_example_evol} shows the evolution of this quantity for the fiducial model. It can be seen that the dust radius expands in time and closely follows the evolution of the gas radius.


To reinforce that the expansion of the dust radius is due to viscosity, \autoref{fig:rdisc_example_evol} also shows the result of the evolution when we do not allow the gas to viscously evolve, but we still consider radial drift. We can see that the dust disc does not expand with time\footnote{The disc undergoes a small expansion at the very beginning of the simulation, despite the fact that the dust velocity is direct inwards at all times. This is not a bug: in a \textit{mathematical} sense the radius of a disc can get larger even if the velocity is always inwards. This is however only a small effect and it is not \textit{physically} important.}, proving that viscosity is the driver of the disc expansion. 

The behaviour of the dust radius is apparently counter-intuitive: one might expect that radial drift causes the discs to simply shrink with time as the grains move closer to the star. We dedicate appendix \ref{sec:mass_evol_appendix} to explaining why instead the disc expands. In a nutshell, radial drift is a victim of its own success: by promoting a rapid inspiral, it removes the fastest-drifting dust, leaving behind relatively well-coupled grains which follow the viscous evolution of the gas.

Finally, we note that at any given time the dust radius is bigger than the gas radius, despite the existence of a sharp outer edge in the dust distribution that we have highlighted in the previous section. This is because the dust has a slightly shallower surface density profile, as can be seen in the top panel of \autoref{fig:summary_0.001_0.1_10}: the dust-to-gas ratio increases towards large radii, as expected in the drift dominated regime (see discussion in \citealt{Birnstiel2012}). This effect is more important in determining the relative dust and gas mass radii than the sharp edge in the dust distribution. Indeed, we find that there is less than 1 per cent of the total gas mass beyond the dust outer edge.

We conclude that the gas viscous spreading influences also the dust and leads to the dust disc becoming larger with time.


\subsection{Time evolution of the flux radius}
\label{sec:example_flux}

\begin{figure}
\includegraphics[width=0.45\textwidth]{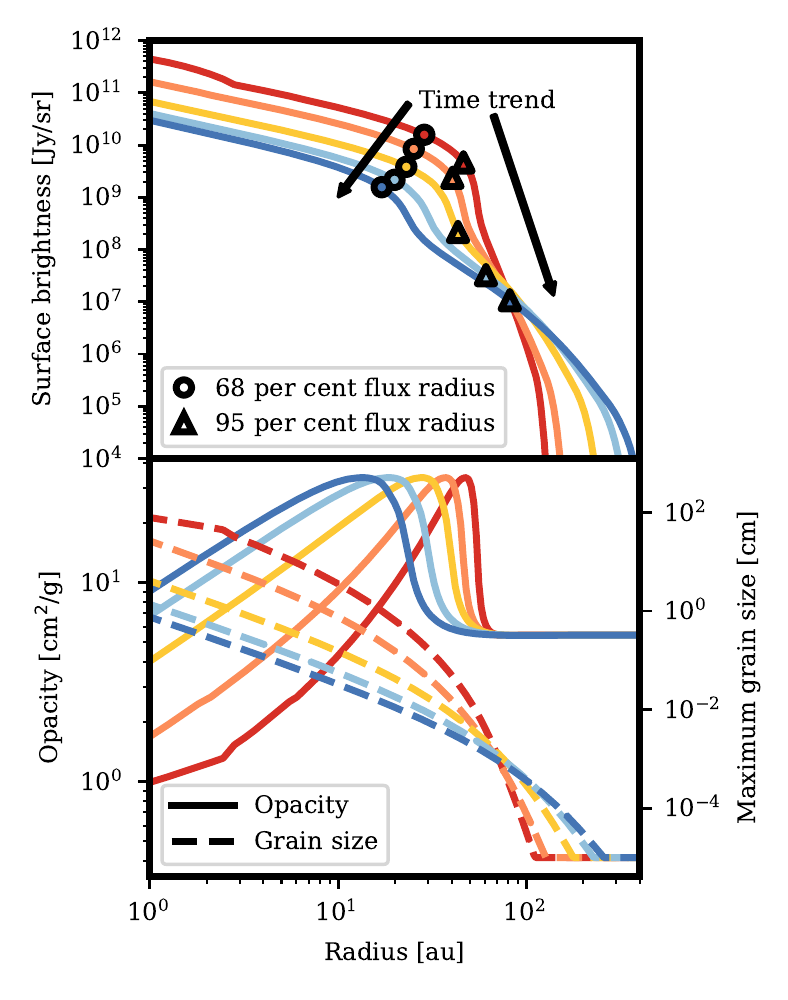}
\caption{\textbf{Top panel:} the surface brightness at different times of the evolution of the disc (0.1, 0.3, 1, 2 and 3 Myr), showing a sharp drop close to the 68 per cent flux radii. As time passes the 68 per cent flux radii shrink, while the 95 per cent expand. \textbf{Bottom panel:} the dust opacity (solid lines) and maximum grain size (dashed lines) at different times (same as in the top panel). The location of the abrupt change in the surface brightness in the top panel corresponds to the location of the peak in the opacity.}
\label{fig:sb_times}
\end{figure}


We now consider the quantity that can be measured by observations, the flux radius. To understand its behaviour, we need to study the surface brightness. The top panel of \autoref{fig:sb_times} shows the surface brightness at 850 $\mu$m of the fiducial model at different times. We note that the surface brightness is composed of two smoothly varying regions, connected by a small region over which the surface brightness varies by a factor of $\sim$ 10. This abrupt variation in surface brightness corresponds to the abrupt change in dust opacity for a grain size of $\sim$ {0.02 cm} that we have called ``opacity cliff''. This is shown by the bottom panel of \autoref{fig:sb_times} in which we plot the dust opacity (solid lines) as a function of radius; for reference we show again also the maximum grain size (dashed lines) already plotted in \autoref{fig:summary_0.001_0.1_10}.

This particular shape of the surface brightness implies that most of the sub-mm flux is coming from a relatively small (few tens of au) region where the grains are larger than the opacity cliff (see \autoref{fig:opacity}) and the surface brightness is therefore relatively high\footnote{Note that the surface brightness always \textit{increases} going towards to the star, even if the opacity \textit{decreases}. This is because the decrease in opacity is more than offset by the higher surface density and temperature close to the star.}. The evolution of this opacity cliff radius does not trace how the mass is evolving, but rather traces the processes controlling grain growth.

Given that most of the flux comes from the region where the opacity is above the cliff, it is not surprising that the cliff radius can be traced quite well by a radius definition based on a given fraction of the disc flux. This is shown by the locations of the dots in the left panel of \autoref{fig:sb_times}, which correspond to the locations of the 68 per cent flux radii.

While in the previous section we have shown that the mass radius increases with time, \autoref{fig:sb_times} show that the opacity cliff moves towards the star with time as a consequence of the grain size becoming smaller at each radius. Therefore, in contrast to the evolution of the mass radius, the 68 per cent flux radii shrink with time. It follows that the 68 percent flux radius can be much smaller than the mass radius: for example, while the 68 per cent flux radius is 20 au at 3 Myr, the mass radius is $\sim$ 100 au (see \autoref{fig:rdisc_example_evol}).

The different time evolution of the flux and mass radii implies that there must be a significant fraction of the dust mass hidden beyond the flux radius. Indeed, there is observational evidence that proto-planetary discs are larger than what is inferred from the sub-mm continuum in alternative tracers: for example in bright molecular emission lines \citep{2005A&A...443..945P,2007A&A...469..213I,2009A&A...501..269P,2012ApJ...744..162A,AnsdellLupusII}, and in a few cases in scattered light observations \citep{2000ApJ...544..895G,2002ApJ...566..409W}.

The small dust in the outer part of the disc has a low surface brightness, but will still contribute somewhat to the disc sub-mm flux. To recover the result that the disc gets larger with time, we need to consider radii definitions based on higher fractions of the total disc flux than the 68 per cent one. For this reason in \autoref{fig:sb_times} we indicate with the triangles the location of the 95 per cent flux radii\footnote{{Empirical tests have shown that using such high fractions of the total disc flux is necessary, even if it has the disadvantage of requiring observations with high signal-to-noise (exceeding that required on the total flux by at least
a factor 100). Lower fractions, for example 80 or 90 per cent, are not enough to recover that the disc expands with time, at least not for all the cases we explore in \autoref{sec:flux}}.}. It can be seen that, after a short initial shrinking phase, these increase with time, tracking the mass distribution. The 95 per cent flux radii trace a faint part of the disc; we will explore in section \ref{sec:limited} the impact of the finite telescope sensitivity on these measurements.

We conclude that the prediction of viscous models is that the dust flux radius increases with time, but only when considering relatively high fractions (95 per cent) of the disc flux. This is a consequence of the small dust opacity (and therefore surface brightness) in the outer part of the disc. In contrast, other definitions like the 68 per cent flux radius trace where the grains are large, rather than the physical extent of the disc.


\section{Dependence on system parameters - evolution of the mass radius}

\label{sec:mass_radius_all}

\subsection{Sensitivity to initial disc radius}

\begin{figure}
\includegraphics[width=\columnwidth]{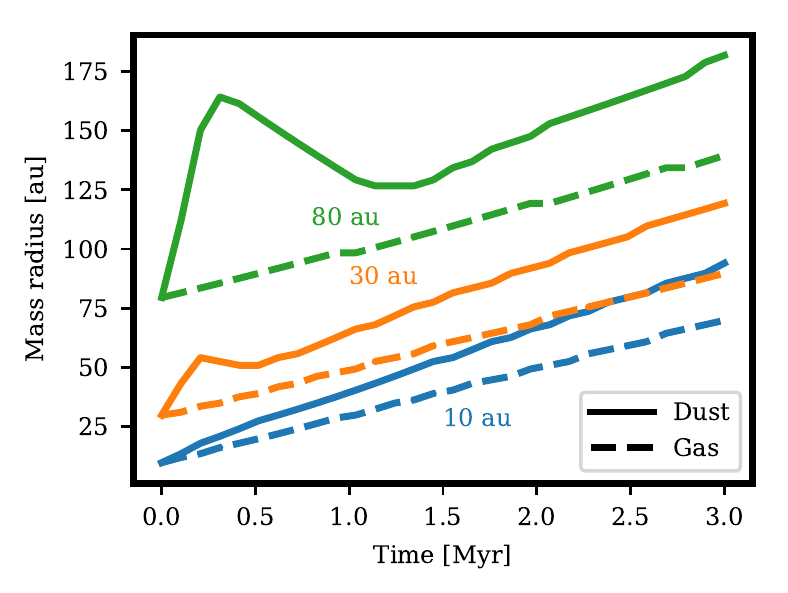}
\caption{Evolution of the dust and gas mass radius for the fiducial case $\alpha=10^{-3}$ for different disc initial radii.}
\label{fig:r_t_0.001}
\end{figure}

For the fiducial case of $\alpha=10^{-3}$, we vary the initial disc radius and study how the evolution of the mass radius changes. We plot the time evolution of the mass radius for different initial disc radii in \autoref{fig:r_t_0.001}. The dust disc always tends to expand, following the expansion of the gas disc; as in the previous case, the dust radius is always larger than the gas radius. For the largest disc we consider (80 au), there is a short-lived shrinking phase. This phase is due to radial drift, which here is more effective in comparison to viscosity due to the longer viscous time scale (4 Myr). The shrinking phase begins after $\sim$ 0.5 Myr because the grains take some time to grow from the initial, sub-$\mu$m sizes up to the limit imposed by radial drift. Note however that the shrinking phase is rather short-lived: by depleting the dust grains, radial drift also causes the grains to become much smaller (see \autoref{eq:a_drift}). In this way the dust grains become coupled to the gas and the dust disc expands again. As we have seen before, radial drift is a victim of its own success, quickly depleting the large grains that are drifting the fastest.


\subsection{Sensitivity to viscosity}

\label{sec:sensitivity_viscosity}

\begin{figure}
\includegraphics[width=\columnwidth]{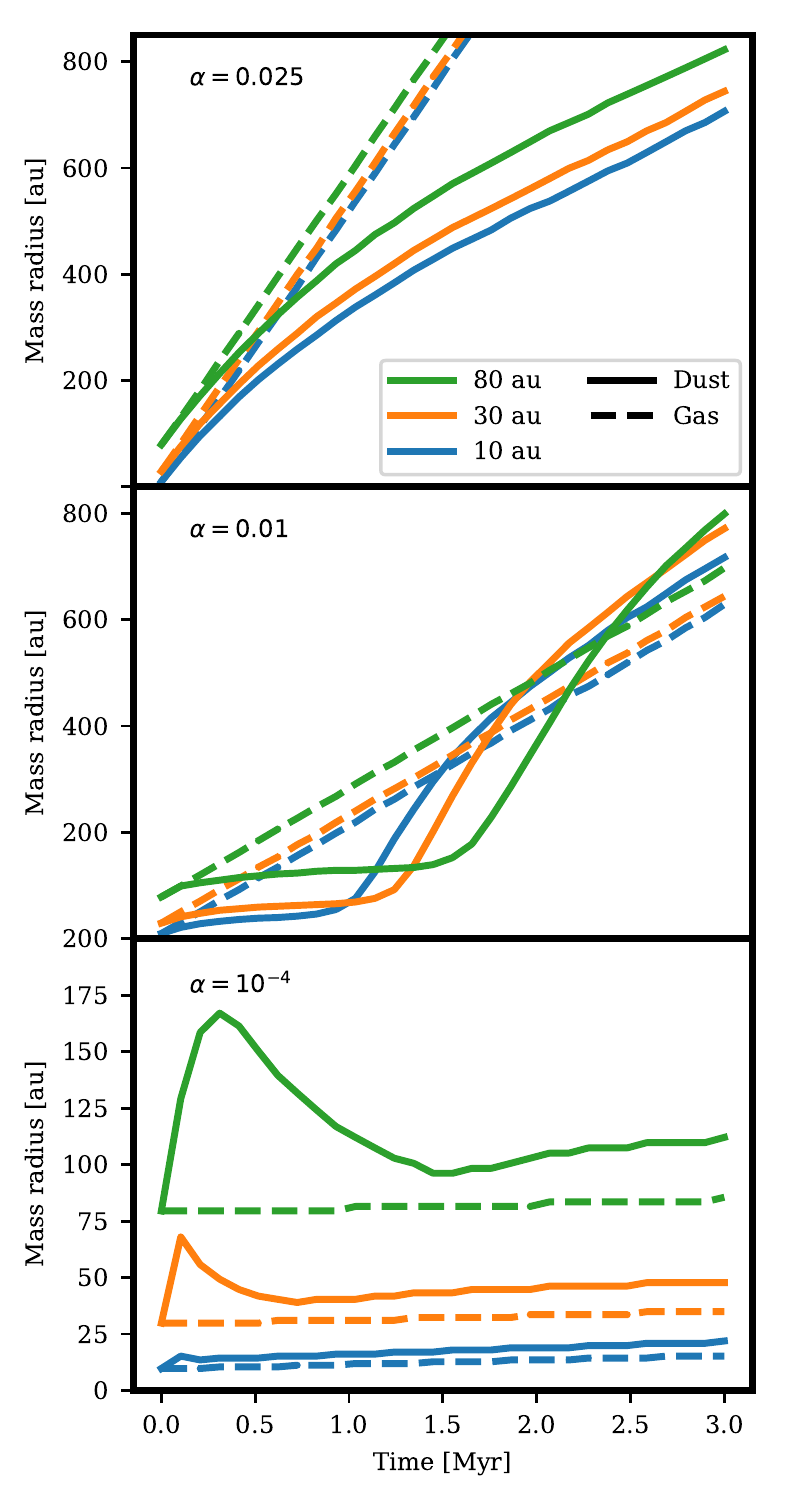}
\caption{Evolution of the dust and gas mass radius for different values of the viscous $\alpha$ parameter and different initial radii. The dust disc expands rapidly with a high value of $\alpha$, whereas the disc radius remains roughly constant if the viscosity is low. Note the different scales on the y axis.}
\label{fig:r_t_alphas}
\end{figure}

\begin{figure}
\includegraphics[width=\columnwidth]{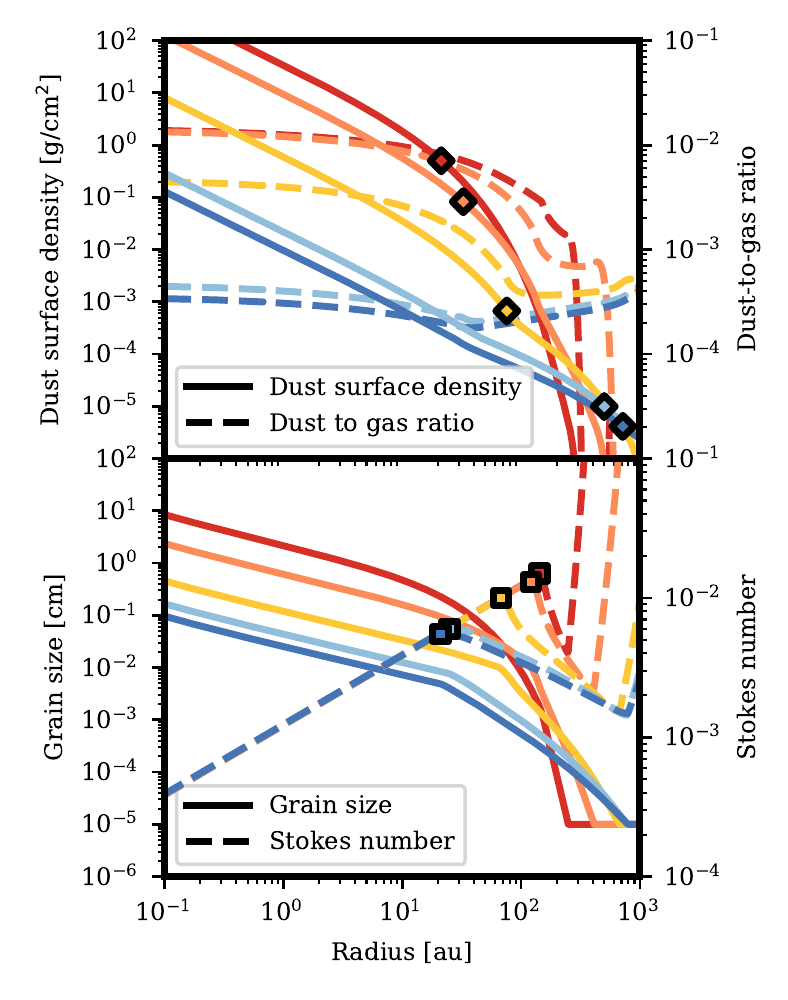}
\caption{Like \autoref{fig:summary_0.001_0.1_10}, but for $\alpha=10^{-2}$. The quantities are plotted at different times (0.1, 0.3, 1, 2 and 3 Myr) as a function of radius. The diamonds in the top panel denote the mass radius at each timestep and the squares in the bottom panel the transition between drift dominated and fragmentation dominated regimes. At the beginning of the simulation most of the disc is in the fragmentation dominated regime, while after $\sim$ 1 Myr of evolution the transition radius between fragmentation and drift dominated has moved inside the dust mass radius.}
\label{fig:amax_sigmas_0.01_0.1_10}
\end{figure}

We show in \autoref{fig:r_t_alphas} the time evolution of the dust and gas mass radius for different values of $\alpha$. In general, high values of $\alpha$ (the $10^{-3}$ case considered previously, and the new cases $\alpha=0.025$ and 0.01 we show here) make the dust spread, whereas a lower value of $10^{-4}$ leads to the dust disc staying roughly constant in radius. 

We will not comment further on the lowest viscosity case; the low spreading rate is simply a consequence of the low amount of viscosity. Even in the gas, viscous spreading is small: the gas mass radius of the 10 au disc does not even double throughout the simulation, as expected since the viscous timescale in this case is 5 Myr.

The case with a viscosity of $\alpha=0.01$ is characterised by two phases of expansion at different rates. Understanding this behaviour necessitates a more detailed look because in this model grain growth proceeds in a qualitatively different way from the fiducial model we have illustrated in section \ref{sec:example_general}. We show in \autoref{fig:amax_sigmas_0.01_0.1_10} the evolution of different quantities in the disc as a function of time and radius. In the top panel, which shows the dust surface density and dust-to-gas ratio, we have marked with the diamonds the dust mass radius at each timestep, whereas in the bottom panel, which shows the grain size and Stokes number, we have marked with the squares the transition between fragmentation and drift dominated regimes. The plot illustrates that at the beginning of the simulation most of the disc is in the fragmentation dominated regime; as the disc depletes however the limit imposed by radial drift becomes more stringent, and after $\sim$ 1 Myr the dust mass radius becomes bigger than the transition radius between being fragmentation and drift dominated. Most of the disc is now in the drift dominated regime and the Stokes numbers of the dust grains have decreased significantly.

Armed with this knowledge, we can now interpret more in detail the evolution of the disc radius that we showed in \autoref{fig:r_t_alphas}. The phase of very rapid expansion around 1 Myr is due to the switch from fragmentation dominated regime to the drift dominated regime previously illustrated. The smaller grain sizes imposed by the drift regime make the dust well coupled, and we find that after the switch the grains move with the gas. Note that, as for the lower viscosity cases, after the switch the dust mass radius is larger than the gas mass radius.

For the highest viscosity case with $\alpha=0.025$, the evolution is rather simple: the radius undergoes a simple, approximately linear expansion. In this case the fragmentation dominated regime is always dominant and the grains are well coupled to the gas at any radius. As a result the evolution of the dust mass radius simply follows the behaviour of the gas. With time, the expansion levels off because the rapid expansion of the disc and accretion of the gas onto the star (promoted by the high viscosity) decrease the gas surface density so significantly that even the smallest grain size we enforce (0.1 $\mu$m) is only partially coupled to the gas.


The difference in viscosity between $\alpha=0.01$ and $\alpha=0.025$ is quite small, but the in-depth look we gave explains why this difference is significant: increasing $\alpha$ simultaneously increases the gas viscous velocity and reduces the dust Stokes number (due to increased fragmentation).


Finally, it is worth noting that the disc radius after a given time is not necessarily a monotonic function of the initial radius; an initially smaller disc can ``overtake'' an initially larger one. This is a consequence of the faster evolution timescales of smaller discs: the grains become well coupled to the gas at earlier times, which is also when the dust radius expands significantly to reach the gas value.

To summarise, while not as straightforward as the evolution of the gas mass radius, overall the amount of viscous spreading in the dust mass radius behaves quite naturally: higher values of the viscosity lead to higher amounts of spreading.

\section{Dependence on system parameters - evolution of the flux radius}

\label{sec:flux}

\begin{figure*}

\parbox{\columnwidth}{
\includegraphics[width=\columnwidth]{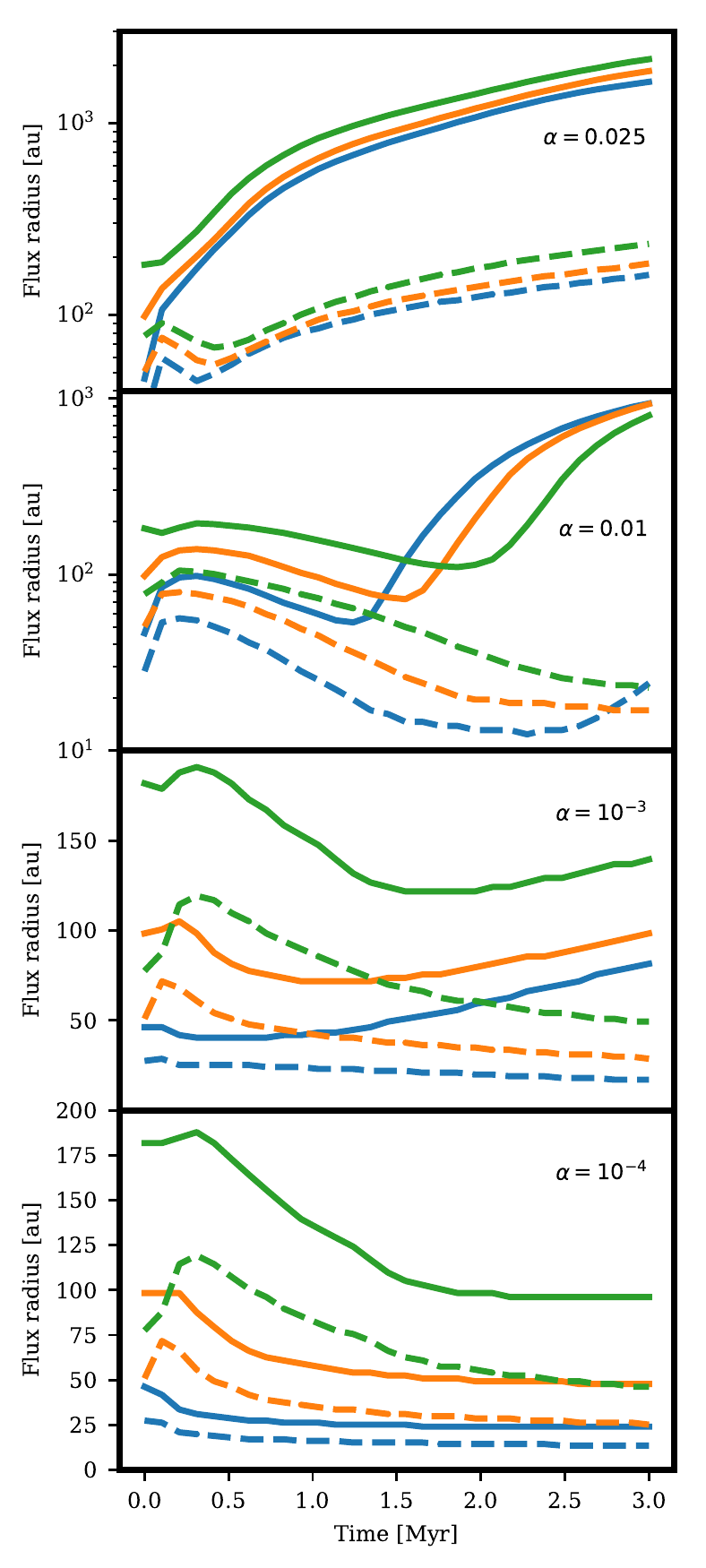}
\centering
\includegraphics[]{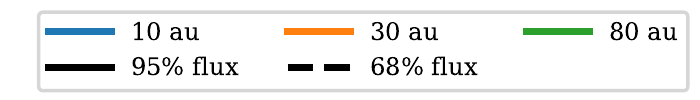}
}
\parbox{\columnwidth}{
\includegraphics[width=\columnwidth]{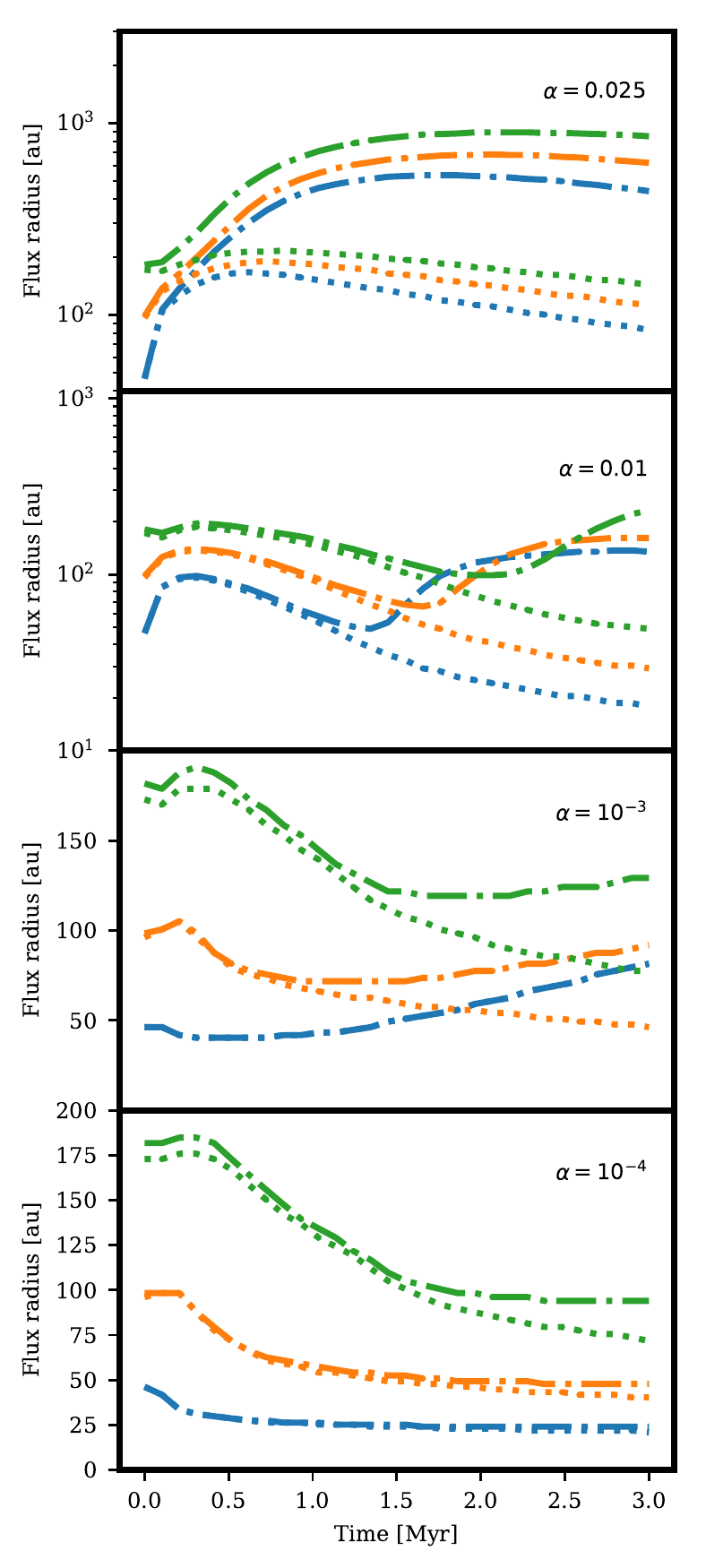}
\centering
\includegraphics[]{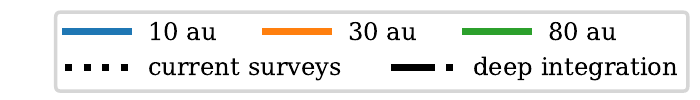}
}

\caption{\textbf{Left panel}: Evolution of the flux radius with time for different viscosities and initial radii. Note the different scales on the y axis. We plot both the 95 per cent radius and the 68 per cent radius since they have different qualitative behaviours: the first expands while the second shrinks. \textbf{Right panel}: Evolution of the 95 per cent flux radius with time when taking into account the finite surface brightness sensitivity of real observations. The linestyles are dotted for existing ALMA surveys (sensitivity of $6 \times 10^{7}$ Jy/sr) and dotted-dashed for a deep integration (sensitivity of $10^6$ Jy/sr). Current surveys are not deep enough to detect viscous spreading, which will require very deep ALMA integrations.}
\label{fig:r_t_flux}
\end{figure*}


The left panel of \autoref{fig:r_t_flux} shows our results for the flux radius, including both the 68 per cent and the 95 per cent radii {(for an easier comparison with the mass radii, we plot these results again side by side with the mass radius evolution in figure \ref{fig:r_t_mass_flux_comparison})}. 
Qualitatively, the radius evolution follows the same features we have described in section \ref{sec:example_flux} and \ref{sec:mass_radius_all}: the 68 per cent flux radii shrink with time while the 95 per cent expand. The rate of expansion of the 95 per cent flux radius increases with $\alpha$, in the same way as the expansion of the mass radius. There are however quantitative differences. The most important difference is that most discs experience an initial shrinking phase, even for the 95 per cent flux radii. We already highlighted this feature for the mass radius evolution, but the flux radius is more affected because of the changes in opacity. The shrinking of the radii is due to the large grains (which dominate the opacity over the smaller grains in the outer part of the disc) rapidly drifting onto the star.

The discs with $\alpha=10^{-4}$ never experience a growing phase after the initial shrinking; their radius remains constant and therefore these discs always remain relatively small. For the other cases instead the 95 per cent radius (solid lines) grow again after the initial short shrinking phase. This expansion is relatively mild for the discs with an intermediate $\alpha=10^{-3}$, which on these timescales attain radii ranging from 50 to 150 au. On the other hand, the dust discs can grow to hundreds of au if the viscosity is high ($\alpha \geq 10^{-2}$).


For the highest viscosity case, there is no initial radius shrinking phase since in those cases the grains are always well coupled to the gas. In addition, even the 68 per cent flux radius simply increases with time. This is because in this case, due to the smaller grain sizes induced by the high turbulence, most grains are smaller than the opacity cliff: we find that soon after the initial conditions only the innermost 10 au of the disc are above the cliff, and this region shrinks further with time. Therefore, the flux is dominated by the emission from small grains and not by large ones (relative to the opacity cliff). {Towards the end of the simulation (after $\sim$2.5 Myr), a similar effect happens also for the 10au disc with $\alpha=10^{-2}$; the 68 percent radius starts increasing rather than decreasing.}

Summarising, viscous spreading is observable also in the \textit{dust} continuum emission, with rates that increase with the value of $\alpha$. This requires however employing a definition of disc radius based on a high fraction (e.g. 95 per cent) of the total flux. Otherwise, using alternative definitions based on smaller fractions of the flux (e.g. 68 per cent), the disc radii in most cases shrink with time as they measure where the grains are larger than the opacity cliff, rather than tracking the mass radius.

\section{Observational consequences}

\label{sec:discussion}

In the previous sections we have shown that, as a consequence of viscous spreading, the dust mass radius expands with time. We stress that these results do not contradict previous investigations \citep{Birnstiel2014} that concluded that the dust disc has a sharp outer edge \textit{at any given time}. In this paper we have instead characterised the evolution of the dust radius \textit{with time}, showing that it tracks the motion of the gas.

We have also highlighted that models of grain growth predict that the sub-mm flux is dominated by a bright central region of the disc where the grains are large enough to have a significant opacity, while additional dust mass can be hidden in the faint outer part of the disc as small grains. A sharp drop in surface brightness (see top panel of \autoref{fig:sb_times}) clearly separates these two regions. The bright inner region shrinks with time, while the faint outer one expands, as tracked respectively by the 68 and 95 per cent flux radii. A crucial question is whether observations are sensitive enough to detect the faint outer region; if not, the observed disc radii would shrink even if discs are getting larger. We dedicate section \ref{sec:limited} to answer this question.

\subsection{Is it possible to observe viscous spreading in the dust?}

\label{sec:limited}

\subsubsection{Are current surveys deep enough?}


In this section we study whether observations are sensitive enough to recover the faint outer part of the disc and therefore detect viscous spreading. Interferometers like ALMA are sensitive only to emission above a given surface brightness. To model the response of the interferometer, we thus discard regions of the disc where the surface brightness is below a given threshold. We then reapply the definitions of observed disc radius of section \ref{sec:example_flux} and \ref{sec:flux} to the resulting surface brightness. To set the threshold, we consider a representative value of the current ALMA surveys. The typical angular resolution is 0.3 arcsec and the integration time a couple of minutes \citep{BarenfeldFirst,Pascucci2016,AnsdellLupusI,Cox2017,Cieza2018}. With these numbers, the ALMA sensitivity calculator\footnote{\url{https://almascience.eso.org/proposing/sensitivity-calculator}} reports a rms noise of 0.15 mJy/beam in band 7 at 850 $\mu$m, which corresponds to $6 \times 10^7$ Jy/sr (or equivalently 1.5 mJy/arcsec$^2$). Note that this exercise is formally independent of the distance from the disc because surface brightness does not depend on distance. The distance however still matters because measuring radii requires enough angular resolution to resolve the discs. At the typical distance of 140 pc from the most studied star forming regions, a resolution of 0.3 arcsec corresponds to $\sim$ 20 au in radius, which should in principle be sufficient for most of the cases we consider here.

In this section we focus on the 95 per cent flux radius since we have shown that the 68 per cent flux radius always shrinks. Measuring viscous spreading therefore requires studying the former. For completeness, we report that the 68 per cent flux radii is in most cases unaffected by the finite surface brightness sensitivity. The only exceptions are for $\alpha=0.025$ because, due to rapid expansion of the disc, the mass is spread over a very large emitting area, resulting in a low surface brightness.

The 95 per cent flux radii, taking into account the sensitivity of current surveys as explained at the beginning of this section, are plotted as the dotted lines in the right panel of \autoref{fig:r_t_flux}. The plot clearly shows that the current ALMA surveys are not deep enough to detect viscous spreading: all the observed disc radii shrink with time. Inspection of \autoref{fig:sb_times} confirms that this cut in surface brightness is not able to recover the parts of the disc emitting beyond the opacity cliff. Therefore, surveys significantly deeper than the ones currently being performed are needed to detect viscous spreading.



\subsubsection{Prospects for deeper surveys}

We repeated this analysis with a deeper threshold to understand whether ALMA has the potential to uncover the low surface brightness part of the disc. Note that, since for an interferometer the sensitivity per beam {does not depend on the resolution}, degrading the angular resolution enhances significantly the surface brightness sensitivity. The requirement to resolve the discs poses limits on how much the resolution can be degraded. Here we consider a surface brightness sensitivity of $10^6$ Jy/sr, which corresponds to an on-source integration time of 1 hr for a beam of 1 arcsec (resulting from the most compact configuration C43-1) or to an integration time of 5 hr for a resolution of 0.67 arcsec (corresponding to configuration C43-2). Especially the latter, corresponding to a resolution in radius of 50 au, should be adequate in most of the cases we present here. We do not attempt to perform more complicated modelling in this paper because our analysis shows that the limiting factor in detecting viscous spreading through observations of the sub-mm continuum is sensitivity, and not angular resolution.

The right panel of \autoref{fig:r_t_flux} shows the results of this exercise as the dotted-dashed lines (again only for the 95 per cent radius). For the cases with $\alpha=10^{-3}$ and $\alpha=10^{-4}$, we recover correctly the theoretical values with infinite sensitivity. For the $\alpha=0.01$ case instead, the observed disc radius is never bigger than $\sim$ 200 au, even if in the left panel of \autoref{fig:r_t_flux} we have shown that, with infinite sensitivity, the disc radius would be several hundreds of au. For the very high viscosity case of $\alpha=0.025$, the top panel shows that we are able to recover large values of hundreds of au. We have already highlighted in section \ref{sec:evol_time_mass} how the apparently small variation between $\alpha=0.01$ and $\alpha=0.025$ is significant and here we find the same: with $\alpha=0.01$ the disc still loses a significant amount of dust onto the star due to radial drift. Combined with the significant expansion of the disc, this lowers considerably the surface brightness of the disc, so that a large part of the emission goes below the detection threshold. On the contrary, in the model with the highest viscosity most of the dust is retained and the disc surface brightness is higher, although even in this case it is at the limit of detection (see for example how the flux radius of the 10 au disc slightly shrinks after 2 Myr).

Given the time evolution of these radii, is it possible to measure viscous spreading? A direct detection would be possible, but challenging. Discounting the $\alpha=10^{-4}$ case (see section \ref{sec:sensitivity_viscosity}), broadly speaking the other discs experience an expanding phase. The main challenge is the existence of an initial shrinking phase, a particularly acute problem for the discs with $\alpha=0.01$ and large initial radii. This shrinking phase corresponds to the phase in which the disc is in the fragmentation dominated regime. In the highest viscosity case of $\alpha=0.025$, the flux radii rapidly saturate to a value of several hundreds of au; while it might not be possible to detect an expansion in this case, such large values of the disc radius would be an indirect evidence of very high values of the viscosity.


As another indirect constraint on viscosity, we also note that the models with values of $\alpha \gtrsim 10^{-3}$ are the only ones in which the disc radii are larger than $\sim$ 100-150 au after a few Myr of evolution. The existence of large discs might thus point to values of the viscosity $\alpha \geq 10^{-3}$ (though see section \ref{sec:caveats} for possible caveats).

In summary, current surveys lack enough sensitivity to detect viscous spreading. Significantly deeper surveys would be needed, although a direct detection of viscous spreading would still be challenging even for ALMA.



%
%
%
%
%

\subsection{Comparison with current sub-mm observations}

Even if current surveys lack the sensitivity to detect viscous spreading, we can still investigate if the current observations support the prediction made in this paper that the 68 per cent flux radii should shrink with time. There are currently four star forming regions with published dust disc radii: Ophiuchus \citep{Cox2017,Cieza2018}, Taurus \citep{Tripathi}, Lupus \citep{Tazzari2017,Andrews2018} and Upper Sco \citep{BarenfeldSizes}. Unfortunately, a straightforward comparison between them is precluded by the fact that the data has been modelled with different approaches. In Ophiuchus the reported disc radii have been derived only by fitting Gaussian profiles to the surface brightness, while in the other cases the disc radii have been measured by fitting power-law profiles. Even inside this broad category, there are still differences that prevent a one-to-one comparison: \citet{Tazzari2017} used a viscous self-similar solution (with an exponential tapering), \citet{Tripathi} and \citet{Andrews2018} a Nuker profile (effectively a broken power-law, with a steep power-law tapering), and \citet{BarenfeldSizes} a truncated power-law. \citet{Tazzari2017} reported that the discs in Lupus are larger for the same brightness than those in Taurus, but this is still a matter of debate since \citet{Tripathi} and \citet{Andrews2018}, using a consistent methodology, did not find any statistically significant difference between the two regions. 

Fits with gaussian profiles are available also for the Lupus region \citep{AnsdellLupusI}. We have compared those results with the radii reported by \citet{Cieza2018} for Ophiuchus, but the two populations are statistically indistinguishable: the p-value computed from a Kolmogorov-Smirnov test is 10 per cent, implying that we cannot reject the hypothesis that the disc radii have been extracted from the same underlying distribution.

Therefore the only possible comparison is between Upper Sco and the combined samples of Taurus and Lupus. The two samples have different ages: the former 5-10 Myr, and the latter 1-3 Myr. Interestingly, \citet{BarenfeldSizes} reported that the discs in Upper Sco are a factor of $\sim3$ more compact than in the other sample, consistent with the predictions that we have presented in this paper. However, the lack of a homogeneous analysis prevents a more quantitative comparison.

In this section we have compared our models with observations in terms only of disc radii; for a more comprehensive study in terms of the observed disc flux-radius correlation, see \citet{Rosotti2019letter}.

\subsection{Disc radius measured from optically thick emission lines}

\begin{figure*}
\includegraphics[width=0.45\textwidth]{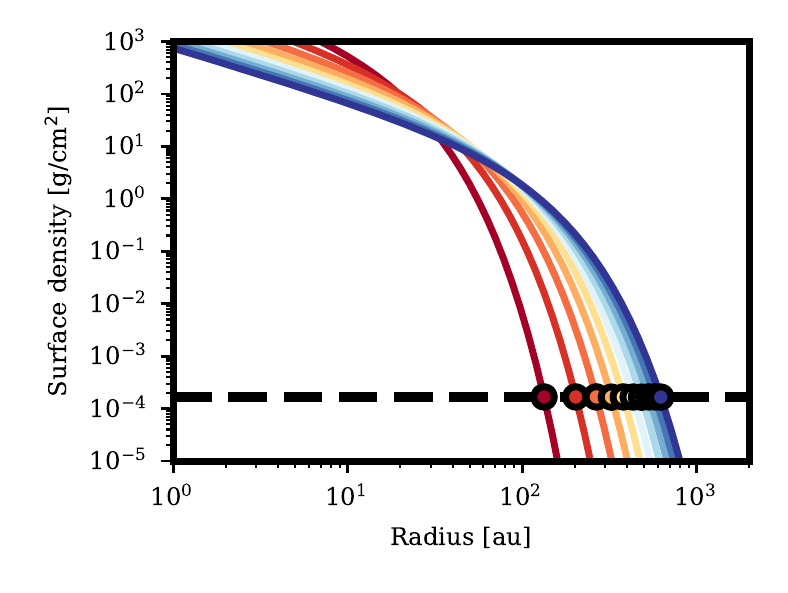}
\includegraphics[width=0.45\textwidth]{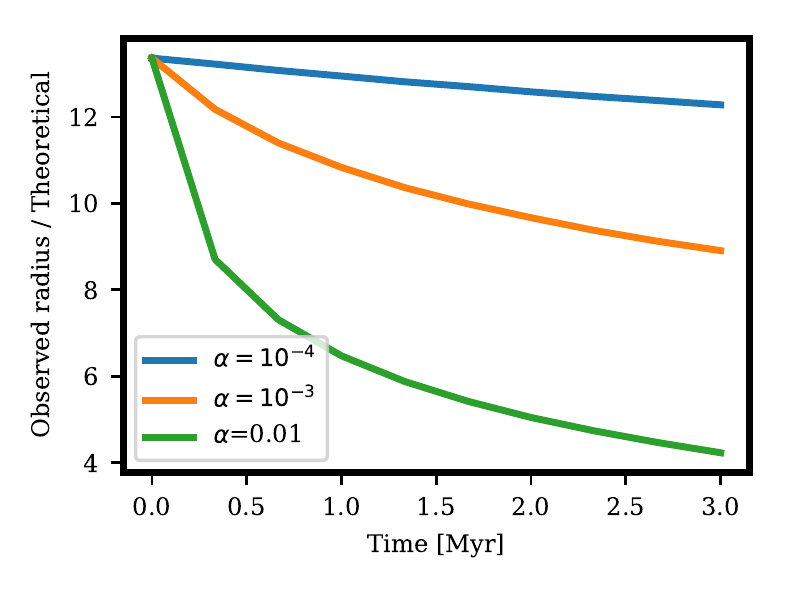}
\caption{\textbf{Left panel}: gas surface density of the fiducial model as a function of time. The horizontal dashed line shows the threshold surface density below which CO dissociates (see text). We mark with dots the radii at which the surface density reaches this threshold, which we define (using an extremely crude assumption) as the radius of the disc. \textbf{Right panel}: the evolution with time of the ratio between the observed radius of the disc, defined as in the left panel, and the gas mass radius of the disc for different values of the viscosity. The ratio is a function of time, preventing its use as a proxy of the disc radius.}
\label{fig:gas}
\end{figure*}

Modelling the gas emission falls outside the scope of this paper, which focuses on the evolution of the dust component of the disc. Nevertheless, in this section we consider if an optically thick gas tracer such as $^{12}$CO can be used to constrain the mechanisms driving disc evolution (as observationally this is relatively easy to access, e.g. \citealt{AnsdellLupusII}). As mentioned in the introduction, an optically thin gas tracer would be ideal, but such observations are extremely time consuming to obtain for a sample of discs.

As a crude assumption, we will assume in what follows that the CO emission traces the part of the disc where the CO column density is higher than the density at which CO self-shielding against the FUV dissociating radiation becomes inefficient \citep{1988ApJ...334..771V}. While crude, this assumption is backed up by thermo-chemical models of discs (see for example figure 9 of \citealt{Cleeves2016}, figure 8 in \citealt{Facchini2017}, the figures in \citealt{2018arXiv180900482M} and \citealt{2019arXiv190306190T}). Following \citet{Facchini2017}, we set this threshold to a column of 10$^{16}$ cm$^{-2}$. This value, corresponding to a total gas surface density of 10$^{20}$ cm$^{-2}$ assuming a standard CO abundance of $10^{-4}$, is slightly higher than the classical value of \citet{1988ApJ...334..771V} due to the different grain size distribution, which affects the UV absorption.

We then consider the gas surface density profiles evolving under the influence of viscosity of the models used in the rest of the paper. We define as radius of the disc the radius at which the surface density falls below the photo-dissociation threshold. We show in the left panel of \autoref{fig:gas} the surface density at different times for the fiducial model and mark with the dots the corresponding radius of the disc. In the right panel we show the corresponding values of the radius as a function of time, normalised to the gas mass radius of the disc, for the fiducial model and other values of the viscosity. If CO was a good tracer of the mass distribution, we would expect these curves to be independent of time, and possibly close to a value of unity. The plot shows instead that this is not the case, with a ratio that can vary significantly depending on the disc parameters.

This issue calls for more detailed modelling than the crude assumption we are taking here. Nevertheless, it demonstrates that CO is a not reliable tracer of the viscous expansion of the disc. It is possible that a detailed modelling of observational data is able to recover correctly the mass radius, but certainly the raw values provided by observations \citep{AnsdellLupusII} cannot be interpreted in a straightforward way.

\subsection{Outlook: on the shape of the surface brightness}
Our models (see for example the top panel of \autoref{fig:sb_times}) predict that the surface brightness should exhibit a sharp drop over a narrow range of radii (see also \citet{Isella2012} for a similar argument). This drop is not caused by the sharp drop in the dust surface density in the outer part of the disc \citep{Birnstiel2014}, but it is an opacity effect. We stress that this prediction is not a particular feature of the grain growth model employed in this paper, as long dust grains are not ``fluffy'' \citep{Kataoka:2014aa}. Rather, it is a consequence of two simple facts: a) a decreasing maximum grain size with radius b) the sharp drop in the sub-mm opacity of dust grains (when $a \lesssim \lambda$) corresponding to the opacity cliff. The first fact is backed by observations of the spectral index variation with radius in proto-planetary discs \citep{2012ApJ...760L..17P,2013A&A...558A..64T,2016A&A...588A..53T,2018ApJ...861...64T} and the second is a general feature of the dependence of dust opacity with grain size for compact dust grains.

We are not aware of discs where such a drop in surface brightness, accompanied by faint emission at larger distances, has been observed. Nevertheless, the fact that some discs are observed to be larger in scattered light than in the sub-mm \citep{2000ApJ...544..895G,2002ApJ...566..409W} seems to corroborate the idea that there might be a population of small grains at large radii. The reason why such a qualitative shape of the surface brightness has not been observed is a lack of surface brightness sensitivity, in the same way as existing surveys lack the sensitivity to detect viscous spreading. Indeed, observations with finite surface brightness sensitivity would likely mistake the drop in surface brightness as the disc outer edge.

While this is speculative at the moment, in the previous section we have indirectly shown that ALMA has the potential, with deep observations, to uncover the outer region of low surface brightness. It will be interesting to see if future, deep observations will detect emission from the disc continuing beyond the currently observed outer edge. This will offer an opportunity to test the assumptions about the opacity and the growth models employed in this paper. While in this paper we have focussed on the surface brightness at a given wavelength, a complementary constraint is also offered by the apparent variation in disc radius when varying the observing wavelength \citep{2018ApJ...861...64T}.


\subsection{Caveats and future directions}

\label{sec:caveats}

\textit{Sub-structure} In this paper we assumed a smooth disc with no substructure. This assumption might seem questionable considering that high-resolution campaigns conducted by ALMA \citep[e.g.,][]{2015ApJ...808L...3A,2016ApJ...820L..40A,2016PhRvL.117y1101I,2017A&A...600A..72F,2018A&A...610A..24F,2018MNRAS.475.5296D,2018A&A...616A..88V,CITau,2018ApJ...869L..41A,2018ApJ...869...17L} and in scattered light \citep{2016A&A...595A.114D,2016A&A...595A.112G,2017ApJ...850...52P,2017ApJ...837..132V,2018ApJ...863...44A} are revealing that many discs possess a high degree of substructure (such as rings and gaps). {It should be borne in mind however that these surveys so far have by necessity targeted only the brightest sources; it remains to be seen how much sub-structure is present in fainter discs that constitute the bulk of the disc population.} 
Our predictions would certainly change quantitatively when considering that the radial substructures imaged in these discs are probably capable of trapping dust. We remark that the expansion of the dust disc is promoted by the viscous expansion of the gas disc at large radii, where the grains are small and relatively well coupled to the gas. Therefore the main results of this paper, that viscous spreading affects also the dust, would still hold as long as the radial traps are sufficiently far from the disc outer edge that the dust grains are free to follow the gas in the outer disc. This is an issue we plan to study in a future paper.

\textit{Constant viscosity} In this paper we employed a constant viscous $\alpha$ over the whole disc. Given the level of uncertainty in the current understanding of the disc accretion mechanisms, we do not think that using more detailed models of the viscosity would be appropriate to this investigation. If $\alpha$ in reality varies with radius, and possibly also with time, the quoted values of $\alpha$ should be regarded as an average across the radial extent of the disc and its lifetime.

\textit{Photo-evaporation} In this paper we did not include processes leading to mass loss in the outer part of the disc, such as external photo-evaporation \citep{1998ApJ...499..758J,2004ApJ...611..360A,2016MNRAS.457.3593F}. While the internal FUV photo-evaporation rates \citep{2009ApJ...690.1539G} are more uncertain due to the lack of hydrodynamical studies, this mechanism could also lead to mass-loss in the outer parts of the disc. This issue is particularly important in the context of this paper because external photo-evaporation removes mass preferentially close to the outer edge of the disc, the same region that is undergoing viscous spreading. A lack of observed disc spreading is therefore not necessarily an evidence that discs do not evolve viscously, but could also be explained as due to the influence of photo-evaporation. Indeed, when considering viscous evolution \citep{Clarke2007}, externally photo-evaporating discs can spread or shrink depending on whether the accretion rate is greater or smaller than the photo-evaporative mass-loss rate. {More in general, in these models we did not include disc dispersal processes. For this reason we restricted ourselves to study the 0-3 Myr time range, comparable to the median disc lifetime \citep[e.g.,][]{Fedele2010}. An extension to older discs, such as those in the Upper Sco star forming region \citep{BarenfeldFirst}, requires including disc dispersal processes in the models since in the region less than 20 percent of the stars still possess a disc \citep{2006ApJ...651L..49C}. Disc dispersal processes are} another issue that we plan to explore in future papers.


\section{Conclusions}

\label{sec:conclusions}

In this paper we have employed models of grain growth and radial drift in proto-planetary discs to study how the disc radius evolves with time. We have investigated both the evolution of how the \textit{mass} is distributed in the disc and, through synthetic surface brightness profiles, the evolution of how the \textit{flux} is distributed, which is relevant for observations. Our main results are as follows:

\begin{itemize}

\item Models of grain growth predict that the dust in the outer parts of discs is small enough to be relatively well coupled to the gas and be entrained in the viscous, outwards flow. While radial drift promotes a rapid inspiral, it is ineffective in overall shrinking dust discs because it quickly removes the fastest drifting grains and becomes a victim of its own success. Therefore, despite radial drift, dust discs get larger with time, as measured by the radius enclosing a given fraction of the total mass, at rates that broadly reflect the efficiency of angular momentum transport in the gas. 

\item We confirm the existence of a sharp outer edge in the dust distribution, as found in the models by \citet{Birnstiel2014}. However, we find that in many cases the dust mass radius (in this paper defined as the 63th percentile of the total mass) is larger than the gas mass radius, a consequence of the different slopes of the gas and dust surface densities.

\item The disc surface brightness is in most cases dominated by the inner part of the disc where the grains are large enough to have a significant sub-mm opacity (above the opacity cliff, see \autoref{fig:opacity}). Therefore, definitions of the disc radius based on the flux, e.g. the 68 per cent flux radius employed in observations, trace the radius inside which the grains are large, rather than the mass radius or the sharp outer edge in the dust distribution. In contrast to the mass radius evolution, these flux radii decrease with time.

\item It is possible to recover observed disc radii that increase with time, if one employs a very high fraction of the total flux (e.g. the 95 per cent flux radius). In addition this requires very deep observations; the surveys currently being performed by ALMA lack enough sensitivity to trace this radius and are instead measuring the part of the disc where the grains are large enough to have significant opacity. While observing viscous spreading will remain a challenge even with these observations, invoking high viscosities ($\gtrsim 10^{-3}$) is the only way to explain large ($\gtrsim 100$ au) discs (unless the depletion of the dust is slowed down by radial traps).



\item Optically thick lines (such as $^{12}$CO) are not a reliable tracer to study viscous spreading since they do not trace the mass radius, but only trace the extent of the optically thick region.

\end{itemize}

\section*{Acknowledgements}

We thank an anonymous referee for a prompt report that significantly improved the clarity of the paper. This work has been supported by the DISCSIM project, grant agreement 341137 funded by the European Research Council under ERC-2013-ADG. This work was performed in part at Aspen Center for Physics, which is supported by National Science Foundation grant PHY-1607611. This work was partially supported by a grant from the Simons Foundation, from the European Union’s Horizon 2020 research and innovation programme under the Marie Skłodowska-Curie grant agreement No 823823 (DUSTBUSTERS) and by the Munich Institute for Astro- and Particle Physics (MIAPP) of the DFG cluster of excellence "Origin and Structure of the Universe". GR acknowledges support from the Netherlands Organisation for Scientific Research (NWO, program number 016.Veni.192.233). MT has been partially supported by the UK Science and Technology research Council (STFC). GL acknowledges support by the project PRIN-INAF 2016 The Cradle of Life - GENESIS-SKA (General Conditions in Early Planetary Systems for the rise of life with SKA). We are grateful to Anna Miotello and Stefano Facchini for interesting discussions about the photo-dissociation of CO.




\bibliographystyle{mnras}
\bibliography{rdisc}

\begin{thebibliography}{}
\makeatletter
\relax
\def\mn@urlcharsother{\let\do\@makeother \do\$\do\&\do\#\do\^\do\_\do\%\do\~}
\def\mn@doi{\begingroup\mn@urlcharsother \@ifnextchar [ {\mn@doi@}
  {\mn@doi@[]}}
\def\mn@doi@[#1]#2{\def\@tempa{#1}\ifx\@tempa\@empty \href
  {http://dx.doi.org/#2} {doi:#2}\else \href {http://dx.doi.org/#2} {#1}\fi
  \endgroup}
\def\mn@eprint#1#2{\mn@eprint@#1:#2::\@nil}
\def\mn@eprint@arXiv#1{\href {http://arxiv.org/abs/#1} {{\tt arXiv:#1}}}
\def\mn@eprint@dblp#1{\href {http://dblp.uni-trier.de/rec/bibtex/#1.xml}
  {dblp:#1}}
\def\mn@eprint@#1:#2:#3:#4\@nil{\def\@tempa {#1}\def\@tempb {#2}\def\@tempc
  {#3}\ifx \@tempc \@empty \let \@tempc \@tempb \let \@tempb \@tempa \fi \ifx
  \@tempb \@empty \def\@tempb {arXiv}\fi \@ifundefined
  {mn@eprint@\@tempb}{\@tempb:\@tempc}{\expandafter \expandafter \csname
  mn@eprint@\@tempb\endcsname \expandafter{\@tempc}}}

\bibitem[\protect\citeauthoryear{{ALMA Partnership} et~al.,}{{ALMA Partnership}
  et~al.}{2015}]{2015ApJ...808L...3A}
{ALMA Partnership} et~al., 2015, \mn@doi [\apj] {10.1088/2041-8205/808/1/L3},
  \href {https://ui.adsabs.harvard.edu/#abs/2015ApJ...808L...3A} {808, L3}

\bibitem[\protect\citeauthoryear{{Adams}, {Hollenbach}, {Laughlin}  \&
  {Gorti}}{{Adams} et~al.}{2004}]{2004ApJ...611..360A}
{Adams} F.~C.,  {Hollenbach} D.,  {Laughlin} G.,   {Gorti} U.,  2004, \mn@doi
  [\apj] {10.1086/421989}, \href
  {https://ui.adsabs.harvard.edu/#abs/2004ApJ...611..360A} {611, 360}

\bibitem[\protect\citeauthoryear{{Andrews} et~al.,}{{Andrews}
  et~al.}{2012}]{2012ApJ...744..162A}
{Andrews} S.~M.,  et~al., 2012, \mn@doi [\apj] {10.1088/0004-637X/744/2/162},
  \href {https://ui.adsabs.harvard.edu/#abs/2012ApJ...744..162A} {744, 162}

\bibitem[\protect\citeauthoryear{{Andrews} et~al.,}{{Andrews}
  et~al.}{2016a}]{Andrews2016}
{Andrews} S.~M.,  et~al., 2016a, \mn@doi [\apj] {10.3847/2041-8205/820/2/L40},
  \href {https://ui.adsabs.harvard.edu/#abs/2016ApJ...820L..40A} {820, L40}

\bibitem[\protect\citeauthoryear{{Andrews} et~al.,}{{Andrews}
  et~al.}{2016b}]{2016ApJ...820L..40A}
{Andrews} S.~M.,  et~al., 2016b, \mn@doi [\apj] {10.3847/2041-8205/820/2/L40},
  \href {https://ui.adsabs.harvard.edu/#abs/2016ApJ...820L..40A} {820, L40}

\bibitem[\protect\citeauthoryear{{Andrews}, {Terrell}, {Tripathi}, {Ansdell},
  {Williams}  \& {Wilner}}{{Andrews} et~al.}{2018a}]{Andrews2018}
{Andrews} S.~M.,  {Terrell} M.,  {Tripathi} A.,  {Ansdell} M.,  {Williams}
  J.~P.,   {Wilner} D.~J.,  2018a, preprint, \href
  {https://ui.adsabs.harvard.edu/#abs/2018arXiv180810510A} {p.
  arXiv:1808.10510} (\mn@eprint {arXiv} {1808.10510})

\bibitem[\protect\citeauthoryear{{Andrews} et~al.,}{{Andrews}
  et~al.}{2018b}]{2018ApJ...869L..41A}
{Andrews} S.~M.,  et~al., 2018b, \mn@doi [\apj] {10.3847/2041-8213/aaf741},
  \href {https://ui.adsabs.harvard.edu/\#abs/2018ApJ...869L..41A} {869, L41}

\bibitem[\protect\citeauthoryear{{Ansdell} et~al.,}{{Ansdell}
  et~al.}{2016}]{AnsdellLupusI}
{Ansdell} M.,  et~al., 2016, \mn@doi [\apj] {10.3847/0004-637X/828/1/46}, \href
  {https://ui.adsabs.harvard.edu/#abs/2016ApJ...828...46A} {828, 46}

\bibitem[\protect\citeauthoryear{{Ansdell}, {Williams}, {Manara}, {Miotello},
  {Facchini}, {van der Marel}, {Testi}  \& {van Dishoeck}}{{Ansdell}
  et~al.}{2017}]{AnsdellSigmaOri}
{Ansdell} M.,  {Williams} J.~P.,  {Manara} C.~F.,  {Miotello} A.,  {Facchini}
  S.,  {van der Marel} N.,  {Testi} L.,   {van Dishoeck} E.~F.,  2017, \mn@doi
  [\aj] {10.3847/1538-3881/aa69c0}, \href
  {https://ui.adsabs.harvard.edu/#abs/2017AJ....153..240A} {153, 240}

\bibitem[\protect\citeauthoryear{{Ansdell} et~al.,}{{Ansdell}
  et~al.}{2018}]{AnsdellLupusII}
{Ansdell} M.,  et~al., 2018, \mn@doi [\apj] {10.3847/1538-4357/aab890}, \href
  {https://ui.adsabs.harvard.edu/#abs/2018ApJ...859...21A} {859, 21}

\bibitem[\protect\citeauthoryear{{Armitage}}{{Armitage}}{2011}]{ArmitageReview}
{Armitage} P.~J.,  2011, \mn@doi [Annual Review of Astronomy and Astrophysics]
  {10.1146/annurev-astro-081710-102521}, \href
  {https://ui.adsabs.harvard.edu/#abs/2011ARA&A..49..195A} {49, 195}

\bibitem[\protect\citeauthoryear{{Avenhaus} et~al.,}{{Avenhaus}
  et~al.}{2018}]{2018ApJ...863...44A}
{Avenhaus} H.,  et~al., 2018, \mn@doi [\apj] {10.3847/1538-4357/aab846}, \href
  {https://ui.adsabs.harvard.edu/#abs/2018ApJ...863...44A} {863, 44}

\bibitem[\protect\citeauthoryear{{Bai} \& {Stone}}{{Bai} \&
  {Stone}}{2013}]{BaiStone2013}
{Bai} X.-N.,  {Stone} J.~M.,  2013, \mn@doi [\apj]
  {10.1088/0004-637X/769/1/76}, \href
  {https://ui.adsabs.harvard.edu/#abs/2013ApJ...769...76B} {769, 76}

\bibitem[\protect\citeauthoryear{{Balbus} \& {Hawley}}{{Balbus} \&
  {Hawley}}{1991}]{BalbusHawley91}
{Balbus} S.~A.,  {Hawley} J.~F.,  1991, \mn@doi [\apj] {10.1086/170270}, \href
  {https://ui.adsabs.harvard.edu/#abs/1991ApJ...376..214B} {376, 214}

\bibitem[\protect\citeauthoryear{{Barenfeld}, {Carpenter}, {Ricci}  \&
  {Isella}}{{Barenfeld} et~al.}{2016}]{BarenfeldFirst}
{Barenfeld} S.~A.,  {Carpenter} J.~M.,  {Ricci} L.,   {Isella} A.,  2016,
  \mn@doi [\apj] {10.3847/0004-637X/827/2/142}, \href
  {https://ui.adsabs.harvard.edu/#abs/2016ApJ...827..142B} {827, 142}

\bibitem[\protect\citeauthoryear{{Barenfeld}, {Carpenter}, {Sargent}, {Isella}
  \& {Ricci}}{{Barenfeld} et~al.}{2017}]{BarenfeldSizes}
{Barenfeld} S.~A.,  {Carpenter} J.~M.,  {Sargent} A.~I.,  {Isella} A.,
  {Ricci} L.,  2017, \mn@doi [\apj] {10.3847/1538-4357/aa989d}, \href
  {https://ui.adsabs.harvard.edu/#abs/2017ApJ...851...85B} {851, 85}

\bibitem[\protect\citeauthoryear{{Bertout}, {Basri}  \& {Bouvier}}{{Bertout}
  et~al.}{1988}]{1988ApJ...330..350B}
{Bertout} C.,  {Basri} G.,   {Bouvier} J.,  1988, \mn@doi [\apj]
  {10.1086/166476}, \href
  {https://ui.adsabs.harvard.edu/#abs/1988ApJ...330..350B} {330, 350}

\bibitem[\protect\citeauthoryear{{Birnstiel} \& {Andrews}}{{Birnstiel} \&
  {Andrews}}{2014}]{Birnstiel2014}
{Birnstiel} T.,  {Andrews} S.~M.,  2014, \mn@doi [\apj]
  {10.1088/0004-637X/780/2/153}, \href
  {https://ui.adsabs.harvard.edu/#abs/2014ApJ...780..153B} {780, 153}

\bibitem[\protect\citeauthoryear{{Birnstiel}, {Dullemond}  \&
  {Brauer}}{{Birnstiel} et~al.}{2009}]{Birnstiel2009}
{Birnstiel} T.,  {Dullemond} C.~P.,   {Brauer} F.,  2009, \mn@doi [\aap]
  {10.1051/0004-6361/200912452}, \href
  {https://ui.adsabs.harvard.edu/#abs/2009A&A...503L...5B} {503, L5}

\bibitem[\protect\citeauthoryear{{Birnstiel}, {Dullemond}  \&
  {Brauer}}{{Birnstiel} et~al.}{2010}]{2010A&A...513A..79B}
{Birnstiel} T.,  {Dullemond} C.~P.,   {Brauer} F.,  2010, \mn@doi [\aap]
  {10.1051/0004-6361/200913731}, \href
  {https://ui.adsabs.harvard.edu/\#abs/2010A&A...513A..79B} {513, A79}

\bibitem[\protect\citeauthoryear{{Birnstiel}, {Klahr}  \&
  {Ercolano}}{{Birnstiel} et~al.}{2012}]{Birnstiel2012}
{Birnstiel} T.,  {Klahr} H.,   {Ercolano} B.,  2012, \mn@doi [\aap]
  {10.1051/0004-6361/201118136}, \href
  {https://ui.adsabs.harvard.edu/#abs/2012A&A...539A.148B} {539, A148}

\bibitem[\protect\citeauthoryear{{Booth}, {Clarke}, {Madhusudhan}  \&
  {Ilee}}{{Booth} et~al.}{2017}]{Booth2017}
{Booth} R.~A.,  {Clarke} C.~J.,  {Madhusudhan} N.,   {Ilee} J.~D.,  2017,
  \mn@doi [\mnras] {10.1093/mnras/stx1103}, \href
  {https://ui.adsabs.harvard.edu/#abs/2017MNRAS.469.3994B} {469, 3994}

\bibitem[\protect\citeauthoryear{{Brauer}, {Dullemond}  \& {Henning}}{{Brauer}
  et~al.}{2008}]{2008A&A...480..859B}
{Brauer} F.,  {Dullemond} C.~P.,   {Henning} T.,  2008, \mn@doi [\aap]
  {10.1051/0004-6361:20077759}, \href
  {https://ui.adsabs.harvard.edu/\#abs/2008A&A...480..859B} {480, 859}

\bibitem[\protect\citeauthoryear{{Carpenter}, {Mamajek}, {Hillenbrand}  \&
  {Meyer}}{{Carpenter} et~al.}{2006}]{2006ApJ...651L..49C}
{Carpenter} J.~M.,  {Mamajek} E.~E.,  {Hillenbrand} L.~A.,   {Meyer} M.~R.,
  2006, \mn@doi [\apj] {10.1086/509121}, \href
  {https://ui.adsabs.harvard.edu/\#abs/2006ApJ...651L..49C} {651, L49}

\bibitem[\protect\citeauthoryear{{Chiang} \& {Goldreich}}{{Chiang} \&
  {Goldreich}}{1997}]{1997ApJ...490..368C}
{Chiang} E.~I.,  {Goldreich} P.,  1997, \mn@doi [\apj] {10.1086/304869}, \href
  {https://ui.adsabs.harvard.edu/#abs/1997ApJ...490..368C} {490, 368}

\bibitem[\protect\citeauthoryear{{Cieza} et~al.,}{{Cieza}
  et~al.}{2018}]{Cieza2018}
{Cieza} L.~A.,  et~al., 2018, preprint, \href
  {http://adsabs.harvard.edu/abs/2018arXiv180908844C} {} (\mn@eprint {arXiv}
  {1809.08844})

\bibitem[\protect\citeauthoryear{{Clarke}}{{Clarke}}{2007}]{Clarke2007}
{Clarke} C.~J.,  2007, \mn@doi [\mnras] {10.1111/j.1365-2966.2007.11547.x},
  \href {https://ui.adsabs.harvard.edu/#abs/2007MNRAS.376.1350C} {376, 1350}

\bibitem[\protect\citeauthoryear{{Clarke} et~al.,}{{Clarke}
  et~al.}{2018}]{CITau}
{Clarke} C.~J.,  et~al., 2018, \mn@doi [\apj] {10.3847/2041-8213/aae36b}, \href
  {https://ui.adsabs.harvard.edu/#abs/2018ApJ...866L...6C} {866, L6}

\bibitem[\protect\citeauthoryear{{Cleeves}, {{\"O}berg}, {Wilner}, {Huang},
  {Loomis}, {Andrews}  \& {Czekala}}{{Cleeves} et~al.}{2016}]{Cleeves2016}
{Cleeves} L.~I.,  {{\"O}berg} K.~I.,  {Wilner} D.~J.,  {Huang} J.,  {Loomis}
  R.~A.,  {Andrews} S.~M.,   {Czekala} I.,  2016, \mn@doi [\apj]
  {10.3847/0004-637X/832/2/110}, \href
  {https://ui.adsabs.harvard.edu/#abs/2016ApJ...832..110C} {832, 110}

\bibitem[\protect\citeauthoryear{{Cox} et~al.,}{{Cox} et~al.}{2017}]{Cox2017}
{Cox} E.~G.,  et~al., 2017, \mn@doi [\apj] {10.3847/1538-4357/aa97e2}, \href
  {https://ui.adsabs.harvard.edu/#abs/2017ApJ...851...83C} {851, 83}

\bibitem[\protect\citeauthoryear{{Dipierro} et~al.,}{{Dipierro}
  et~al.}{2018a}]{2018MNRAS.475.5296D}
{Dipierro} G.,  et~al., 2018a, \mn@doi [\mnras] {10.1093/mnras/sty181}, \href
  {https://ui.adsabs.harvard.edu/#abs/2018MNRAS.475.5296D} {475, 5296}

\bibitem[\protect\citeauthoryear{{Dipierro}, {Laibe}, {Alexander}  \&
  {Hutchison}}{{Dipierro} et~al.}{2018b}]{2018MNRAS.479.4187D}
{Dipierro} G.,  {Laibe} G.,  {Alexander} R.,   {Hutchison} M.,  2018b, \mn@doi
  [\mnras] {10.1093/mnras/sty1701}, \href
  {https://ui.adsabs.harvard.edu/#abs/2018MNRAS.479.4187D} {479, 4187}

\bibitem[\protect\citeauthoryear{{Facchini}, {Clarke}  \& {Bisbas}}{{Facchini}
  et~al.}{2016}]{2016MNRAS.457.3593F}
{Facchini} S.,  {Clarke} C.~J.,   {Bisbas} T.~G.,  2016, \mn@doi [\mnras]
  {10.1093/mnras/stw240}, \href
  {https://ui.adsabs.harvard.edu/#abs/2016MNRAS.457.3593F} {457, 3593}

\bibitem[\protect\citeauthoryear{{Facchini}, {Birnstiel}, {Bruderer}  \& {van
  Dishoeck}}{{Facchini} et~al.}{2017}]{Facchini2017}
{Facchini} S.,  {Birnstiel} T.,  {Bruderer} S.,   {van Dishoeck} E.~F.,  2017,
  \mn@doi [\aap] {10.1051/0004-6361/201630329}, \href
  {https://ui.adsabs.harvard.edu/#abs/2017A&A...605A..16F} {605, A16}

\bibitem[\protect\citeauthoryear{{Fedele}, {van den Ancker}, {Henning},
  {Jayawardhana}  \& {Oliveira}}{{Fedele} et~al.}{2010}]{Fedele2010}
{Fedele} D.,  {van den Ancker} M.~E.,  {Henning} T.,  {Jayawardhana} R.,
  {Oliveira} J.~M.,  2010, \mn@doi [\aap] {10.1051/0004-6361/200912810}, \href
  {https://ui.adsabs.harvard.edu/\#abs/2010A&A...510A..72F} {510, A72}

\bibitem[\protect\citeauthoryear{{Fedele} et~al.,}{{Fedele}
  et~al.}{2017}]{2017A&A...600A..72F}
{Fedele} D.,  et~al., 2017, \mn@doi [\aap] {10.1051/0004-6361/201629860}, \href
  {https://ui.adsabs.harvard.edu/#abs/2017A&A...600A..72F} {600, A72}

\bibitem[\protect\citeauthoryear{{Fedele} et~al.,}{{Fedele}
  et~al.}{2018}]{2018A&A...610A..24F}
{Fedele} D.,  et~al., 2018, \mn@doi [\aap] {10.1051/0004-6361/201731978}, \href
  {https://ui.adsabs.harvard.edu/#abs/2018A&A...610A..24F} {610, A24}

\bibitem[\protect\citeauthoryear{{Flaherty}, {Hughes}, {Teague}, {Simon},
  {Andrews}  \& {Wilner}}{{Flaherty} et~al.}{2018}]{2018ApJ...856..117F}
{Flaherty} K.~M.,  {Hughes} A.~M.,  {Teague} R.,  {Simon} J.~B.,  {Andrews}
  S.~M.,   {Wilner} D.~J.,  2018, \mn@doi [\apj] {10.3847/1538-4357/aab615},
  \href {https://ui.adsabs.harvard.edu/#abs/2018ApJ...856..117F} {856, 117}

\bibitem[\protect\citeauthoryear{{Fromang}, {Latter}, {Lesur}  \&
  {Ogilvie}}{{Fromang} et~al.}{2013}]{Fromang2013}
{Fromang} S.,  {Latter} H.,  {Lesur} G.,   {Ogilvie} G.~I.,  2013, \mn@doi
  [\aap] {10.1051/0004-6361/201220016}, \href
  {https://ui.adsabs.harvard.edu/#abs/2013A&A...552A..71F} {552, A71}

\bibitem[\protect\citeauthoryear{{Garaud}}{{Garaud}}{2007}]{2007ApJ...671.2091G}
{Garaud} P.,  2007, \mn@doi [\apj] {10.1086/523090}, \href
  {https://ui.adsabs.harvard.edu/#abs/2007ApJ...671.2091G} {671, 2091}

\bibitem[\protect\citeauthoryear{{Ginski} et~al.,}{{Ginski}
  et~al.}{2016}]{2016A&A...595A.112G}
{Ginski} C.,  et~al., 2016, \mn@doi [\aap] {10.1051/0004-6361/201629265}, \href
  {https://ui.adsabs.harvard.edu/#abs/2016A&A...595A.112G} {595, A112}

\bibitem[\protect\citeauthoryear{{Gorti} \& {Hollenbach}}{{Gorti} \&
  {Hollenbach}}{2009}]{2009ApJ...690.1539G}
{Gorti} U.,  {Hollenbach} D.,  2009, \mn@doi [\apj]
  {10.1088/0004-637X/690/2/1539}, \href
  {https://ui.adsabs.harvard.edu/#abs/2009ApJ...690.1539G} {690, 1539}

\bibitem[\protect\citeauthoryear{{Grady} et~al.,}{{Grady}
  et~al.}{2000}]{2000ApJ...544..895G}
{Grady} C.~A.,  et~al., 2000, \mn@doi [\apj] {10.1086/317222}, \href
  {https://ui.adsabs.harvard.edu/#abs/2000ApJ...544..895G} {544, 895}

\bibitem[\protect\citeauthoryear{{Hartigan}, {Edwards}  \&
  {Ghandour}}{{Hartigan} et~al.}{1995}]{1995ApJ...452..736H}
{Hartigan} P.,  {Edwards} S.,   {Ghandour} L.,  1995, \mn@doi [\apj]
  {10.1086/176344}, \href
  {https://ui.adsabs.harvard.edu/#abs/1995ApJ...452..736H} {452, 736}

\bibitem[\protect\citeauthoryear{{Hughes}, {Wilner}, {Qi}  \&
  {Hogerheijde}}{{Hughes} et~al.}{2008}]{Hughes2008}
{Hughes} A.~M.,  {Wilner} D.~J.,  {Qi} C.,   {Hogerheijde} M.~R.,  2008,
  \mn@doi [\apj] {10.1086/586730}, \href
  {http://adsabs.harvard.edu/abs/2008ApJ...678.1119H} {678, 1119}

\bibitem[\protect\citeauthoryear{{Isella}, {Testi}, {Natta}, {Neri}, {Wilner}
  \& {Qi}}{{Isella} et~al.}{2007}]{2007A&A...469..213I}
{Isella} A.,  {Testi} L.,  {Natta} A.,  {Neri} R.,  {Wilner} D.,   {Qi} C.,
  2007, \mn@doi [\aap] {10.1051/0004-6361:20077385}, \href
  {https://ui.adsabs.harvard.edu/\#abs/2007A&A...469..213I} {469, 213}

\bibitem[\protect\citeauthoryear{{Isella}, {P{\'e}rez}  \&
  {Carpenter}}{{Isella} et~al.}{2012}]{Isella2012}
{Isella} A.,  {P{\'e}rez} L.~M.,   {Carpenter} J.~M.,  2012, \mn@doi [\apj]
  {10.1088/0004-637X/747/2/136}, \href
  {http://adsabs.harvard.edu/abs/2012ApJ...747..136I} {747, 136}

\bibitem[\protect\citeauthoryear{{Isella} et~al.,}{{Isella}
  et~al.}{2016}]{2016PhRvL.117y1101I}
{Isella} A.,  et~al., 2016, \mn@doi [\prl] {10.1103/PhysRevLett.117.251101},
  \href {https://ui.adsabs.harvard.edu/#abs/2016PhRvL.117y1101I} {117, 251101}

\bibitem[\protect\citeauthoryear{{Johnstone}, {Hollenbach}  \&
  {Bally}}{{Johnstone} et~al.}{1998}]{1998ApJ...499..758J}
{Johnstone} D.,  {Hollenbach} D.,   {Bally} J.,  1998, \mn@doi [\apj]
  {10.1086/305658}, \href
  {https://ui.adsabs.harvard.edu/#abs/1998ApJ...499..758J} {499, 758}

\bibitem[\protect\citeauthoryear{{Kataoka}, {Okuzumi}, {Tanaka}  \&
  {Nomura}}{{Kataoka} et~al.}{2014}]{Kataoka:2014aa}
{Kataoka} A.,  {Okuzumi} S.,  {Tanaka} H.,   {Nomura} H.,  2014, \mn@doi [\aap]
  {10.1051/0004-6361/201323199}, \href
  {http://adsabs.harvard.edu/abs/2014A%26A...568A..42K} {568, A42}

\bibitem[\protect\citeauthoryear{{Laibe} \& {Price}}{{Laibe} \&
  {Price}}{2014}]{LaibePrice2014}
{Laibe} G.,  {Price} D.~J.,  2014, \mn@doi [\mnras] {10.1093/mnras/stu1367},
  \href {https://ui.adsabs.harvard.edu/#abs/2014MNRAS.444.1940L} {444, 1940}

\bibitem[\protect\citeauthoryear{{Lodato}, {Scardoni}, {Manara}  \&
  {Testi}}{{Lodato} et~al.}{2017}]{Lodato2017}
{Lodato} G.,  {Scardoni} C.~E.,  {Manara} C.~F.,   {Testi} L.,  2017, \mn@doi
  [\mnras] {10.1093/mnras/stx2273}, \href
  {https://ui.adsabs.harvard.edu/#abs/2017MNRAS.472.4700L} {472, 4700}

\bibitem[\protect\citeauthoryear{{Long} et~al.,}{{Long}
  et~al.}{2018}]{2018ApJ...869...17L}
{Long} F.,  et~al., 2018, \mn@doi [\apj] {10.3847/1538-4357/aae8e1}, \href
  {https://ui.adsabs.harvard.edu/\#abs/2018ApJ...869...17L} {869, 17}

\bibitem[\protect\citeauthoryear{{Lynden-Bell} \& {Pringle}}{{Lynden-Bell} \&
  {Pringle}}{1974}]{LyndenBellPringle74}
{Lynden-Bell} D.,  {Pringle} J.~E.,  1974, \mn@doi [\mnras]
  {10.1093/mnras/168.3.603}, \href
  {https://ui.adsabs.harvard.edu/#abs/1974MNRAS.168..603L} {168, 603}

\bibitem[\protect\citeauthoryear{{Mathis}, {Rumpl}  \& {Nordsieck}}{{Mathis}
  et~al.}{1977}]{1977ApJ...217..425M}
{Mathis} J.~S.,  {Rumpl} W.,   {Nordsieck} K.~H.,  1977, \mn@doi [\apj]
  {10.1086/155591}, \href
  {https://ui.adsabs.harvard.edu/#abs/1977ApJ...217..425M} {217, 425}

\bibitem[\protect\citeauthoryear{{Miotello}, {Facchini}, {van Dishoeck}  \&
  {Bruderer}}{{Miotello} et~al.}{2018}]{2018arXiv180900482M}
{Miotello} A.,  {Facchini} S.,  {van Dishoeck} E.~F.,   {Bruderer} S.,  2018,
  preprint, \href {http://adsabs.harvard.edu/abs/2018arXiv180900482M} {}
  (\mn@eprint {arXiv} {1809.00482})

\bibitem[\protect\citeauthoryear{{Mulders}, {Pascucci}, {Manara}, {Testi},
  {Herczeg}, {Henning}, {Mohanty}  \& {Lodato}}{{Mulders}
  et~al.}{2017}]{Mulders2017}
{Mulders} G.~D.,  {Pascucci} I.,  {Manara} C.~F.,  {Testi} L.,  {Herczeg}
  G.~J.,  {Henning} T.,  {Mohanty} S.,   {Lodato} G.,  2017, \mn@doi [\apj]
  {10.3847/1538-4357/aa8906}, \href
  {https://ui.adsabs.harvard.edu/#abs/2017ApJ...847...31M} {847, 31}

\bibitem[\protect\citeauthoryear{{Natta} \& {Testi}}{{Natta} \&
  {Testi}}{2004}]{Natta:2004yu}
{Natta} A.,  {Testi} L.,  2004, in {Johnstone} D.,  {Adams} F.~C.,  {Lin}
  D.~N.~C.,  {Neufeeld} D.~A.,   {Ostriker} E.~C.,  eds,  Astronomical Society
  of the Pacific Conference Series Vol. 323, Star Formation in the Interstellar
  Medium: In Honor of David Hollenbach. p.~279

\bibitem[\protect\citeauthoryear{{Natta}, {Testi}, {Calvet}, {Henning},
  {Waters}  \& {Wilner}}{{Natta} et~al.}{2007}]{Natta:2007ye}
{Natta} A.,  {Testi} L.,  {Calvet} N.,  {Henning} T.,  {Waters} R.,   {Wilner}
  D.,  2007, Protostars and Planets V, \href
  {http://adsabs.harvard.edu/abs/2007prpl.conf..767N} {pp 767--781}

\bibitem[\protect\citeauthoryear{{Owen}, {Ercolano}, {Clarke}  \&
  {Alexander}}{{Owen} et~al.}{2010}]{Owen2010}
{Owen} J.~E.,  {Ercolano} B.,  {Clarke} C.~J.,   {Alexander} R.~D.,  2010,
  \mn@doi [\mnras] {10.1111/j.1365-2966.2009.15771.x}, \href
  {https://ui.adsabs.harvard.edu/#abs/2010MNRAS.401.1415O} {401, 1415}

\bibitem[\protect\citeauthoryear{{Pani{\'c}}, {Hogerheijde}, {Wilner}  \&
  {Qi}}{{Pani{\'c}} et~al.}{2009}]{2009A&A...501..269P}
{Pani{\'c}} O.,  {Hogerheijde} M.~R.,  {Wilner} D.,   {Qi} C.,  2009, \mn@doi
  [\aap] {10.1051/0004-6361/200911883}, \href
  {https://ui.adsabs.harvard.edu/#abs/2009A&A...501..269P} {501, 269}

\bibitem[\protect\citeauthoryear{{Pascucci} et~al.,}{{Pascucci}
  et~al.}{2016}]{Pascucci2016}
{Pascucci} I.,  et~al., 2016, \mn@doi [\apj] {10.3847/0004-637X/831/2/125},
  \href {https://ui.adsabs.harvard.edu/#abs/2016ApJ...831..125P} {831, 125}

\bibitem[\protect\citeauthoryear{{P{\'e}rez} et~al.,}{{P{\'e}rez}
  et~al.}{2012}]{2012ApJ...760L..17P}
{P{\'e}rez} L.~M.,  et~al., 2012, \mn@doi [\apj] {10.1088/2041-8205/760/1/L17},
  \href {https://ui.adsabs.harvard.edu/#abs/2012ApJ...760L..17P} {760, L17}

\bibitem[\protect\citeauthoryear{{Pi{\'e}tu}, {Guilloteau}  \&
  {Dutrey}}{{Pi{\'e}tu} et~al.}{2005}]{2005A&A...443..945P}
{Pi{\'e}tu} V.,  {Guilloteau} S.,   {Dutrey} A.,  2005, \mn@doi [\aap]
  {10.1051/0004-6361:20042050}, \href
  {https://ui.adsabs.harvard.edu/\#abs/2005A&A...443..945P} {443, 945}

\bibitem[\protect\citeauthoryear{{Pohl} et~al.,}{{Pohl}
  et~al.}{2017}]{2017ApJ...850...52P}
{Pohl} A.,  et~al., 2017, \mn@doi [\apj] {10.3847/1538-4357/aa94c2}, \href
  {https://ui.adsabs.harvard.edu/#abs/2017ApJ...850...52P} {850, 52}

\bibitem[\protect\citeauthoryear{{Pollack}, {Hollenbach}, {Beckwith},
  {Simonelli}, {Roush}  \& {Fong}}{{Pollack}
  et~al.}{1994}]{1994ApJ...421..615P}
{Pollack} J.~B.,  {Hollenbach} D.,  {Beckwith} S.,  {Simonelli} D.~P.,  {Roush}
  T.,   {Fong} W.,  1994, \mn@doi [\apj] {10.1086/173677}, \href
  {https://ui.adsabs.harvard.edu/#abs/1994ApJ...421..615P} {421, 615}

\bibitem[\protect\citeauthoryear{{Rosotti}, {Clarke}, {Manara}  \&
  {Facchini}}{{Rosotti} et~al.}{2017}]{Rosotti2017}
{Rosotti} G.~P.,  {Clarke} C.~J.,  {Manara} C.~F.,   {Facchini} S.,  2017,
  \mn@doi [\mnras] {10.1093/mnras/stx595}, \href
  {https://ui.adsabs.harvard.edu/#abs/2017MNRAS.468.1631R} {468, 1631}

\bibitem[\protect\citeauthoryear{{Rosotti}, {Booth}, {Tazzari}, {Clarke},
  {Lodato}  \& {Testi}}{{Rosotti} et~al.}{2019}]{Rosotti2019letter}
{Rosotti} G.~P.,  {Booth} R.~A.,  {Tazzari} M.,  {Clarke} C.~J.,  {Lodato} G.,
   {Testi} L.,  2019, \mnras, in press

\bibitem[\protect\citeauthoryear{{Ru{\'\i}z-Rodr{\'\i}guez}
  et~al.,}{{Ru{\'\i}z-Rodr{\'\i}guez} et~al.}{2018}]{2018MNRAS.478.3674R}
{Ru{\'\i}z-Rodr{\'\i}guez} D.,  et~al., 2018, \mn@doi [\mnras]
  {10.1093/mnras/sty1351}, \href
  {https://ui.adsabs.harvard.edu/#abs/2018MNRAS.478.3674R} {478, 3674}

\bibitem[\protect\citeauthoryear{{Shakura} \& {Sunyaev}}{{Shakura} \&
  {Sunyaev}}{1973}]{ShakuraSunyaev1973}
{Shakura} N.~I.,  {Sunyaev} R.~A.,  1973, \aap, \href
  {https://ui.adsabs.harvard.edu/#abs/1973A&A....24..337S} {500, 33}

\bibitem[\protect\citeauthoryear{{Suzuki} \& {Inutsuka}}{{Suzuki} \&
  {Inutsuka}}{2009}]{SuzukiInutsuka2009}
{Suzuki} T.~K.,  {Inutsuka} S.-i.,  2009, \mn@doi [\apj]
  {10.1088/0004-637X/691/1/L49}, \href
  {https://ui.adsabs.harvard.edu/#abs/2009ApJ...691L..49S} {691, L49}

\bibitem[\protect\citeauthoryear{{Takeuchi} \& {Lin}}{{Takeuchi} \&
  {Lin}}{2002}]{TakeuchiLin2002}
{Takeuchi} T.,  {Lin} D.~N.~C.,  2002, \mn@doi [\apj] {10.1086/344437}, \href
  {https://ui.adsabs.harvard.edu/#abs/2002ApJ...581.1344T} {581, 1344}

\bibitem[\protect\citeauthoryear{{Takeuchi}, {Clarke}  \& {Lin}}{{Takeuchi}
  et~al.}{2005}]{TakeuchiClarkeLin2005}
{Takeuchi} T.,  {Clarke} C.~J.,   {Lin} D.~N.~C.,  2005, \mn@doi [\apj]
  {10.1086/430393}, \href
  {https://ui.adsabs.harvard.edu/#abs/2005ApJ...627..286T} {627, 286}

\bibitem[\protect\citeauthoryear{{Tazzari} et~al.,}{{Tazzari}
  et~al.}{2016}]{2016A&A...588A..53T}
{Tazzari} M.,  et~al., 2016, \mn@doi [\aap] {10.1051/0004-6361/201527423},
  \href {https://ui.adsabs.harvard.edu/#abs/2016A&A...588A..53T} {588, A53}

\bibitem[\protect\citeauthoryear{{Tazzari} et~al.,}{{Tazzari}
  et~al.}{2017}]{Tazzari2017}
{Tazzari} M.,  et~al., 2017, \mn@doi [\aap] {10.1051/0004-6361/201730890},
  \href {https://ui.adsabs.harvard.edu/#abs/2017A&A...606A..88T} {606, A88}

\bibitem[\protect\citeauthoryear{{Trapman}, {Facchini}, {Hogerheijde}, {van
  Dishoeck}  \& {Bruderer}}{{Trapman} et~al.}{2019}]{2019arXiv190306190T}
{Trapman} L.,  {Facchini} S.,  {Hogerheijde} M.~R.,  {van Dishoeck} E.~F.,
  {Bruderer} S.,  2019, arXiv e-prints, \href
  {https://ui.adsabs.harvard.edu/\#abs/2019arXiv190306190T} {p.
  arXiv:1903.06190}

\bibitem[\protect\citeauthoryear{{Tripathi}, {Andrews}, {Birnstiel}  \&
  {Wilner}}{{Tripathi} et~al.}{2017}]{Tripathi}
{Tripathi} A.,  {Andrews} S.~M.,  {Birnstiel} T.,   {Wilner} D.~J.,  2017,
  \mn@doi [\apj] {10.3847/1538-4357/aa7c62}, \href
  {https://ui.adsabs.harvard.edu/#abs/2017ApJ...845...44T} {845, 44}

\bibitem[\protect\citeauthoryear{{Tripathi} et~al.,}{{Tripathi}
  et~al.}{2018}]{2018ApJ...861...64T}
{Tripathi} A.,  et~al., 2018, \mn@doi [\apj] {10.3847/1538-4357/aac5d6}, \href
  {https://ui.adsabs.harvard.edu/#abs/2018ApJ...861...64T} {861, 64}

\bibitem[\protect\citeauthoryear{{Trotta}, {Testi}, {Natta}, {Isella}  \&
  {Ricci}}{{Trotta} et~al.}{2013}]{2013A&A...558A..64T}
{Trotta} F.,  {Testi} L.,  {Natta} A.,  {Isella} A.,   {Ricci} L.,  2013,
  \mn@doi [\aap] {10.1051/0004-6361/201321896}, \href
  {https://ui.adsabs.harvard.edu/#abs/2013A&A...558A..64T} {558, A64}

\bibitem[\protect\citeauthoryear{{Turner}, {Fromang}, {Gammie}, {Klahr},
  {Lesur}, {Wardle}  \& {Bai}}{{Turner} et~al.}{2014}]{TurnerReview}
{Turner} N.~J.,  {Fromang} S.,  {Gammie} C.,  {Klahr} H.,  {Lesur} G.,
  {Wardle} M.,   {Bai} X.~N.,  2014, in Protostars and Planets VI. p.~411
  (\mn@eprint {arXiv} {1401.7306}),
  \mn@doi{10.2458/azu_uapress_9780816531240-ch018}

\bibitem[\protect\citeauthoryear{{Weidenschilling}}{{Weidenschilling}}{1977}]{Weidenschilling1977}
{Weidenschilling} S.~J.,  1977, \mn@doi [\mnras] {10.1093/mnras/180.1.57},
  \href {https://ui.adsabs.harvard.edu/#abs/1977MNRAS.180...57W} {180, 57}

\bibitem[\protect\citeauthoryear{{Weinberger} et~al.,}{{Weinberger}
  et~al.}{2002}]{2002ApJ...566..409W}
{Weinberger} A.~J.,  et~al., 2002, \mn@doi [\apj] {10.1086/338076}, \href
  {https://ui.adsabs.harvard.edu/#abs/2002ApJ...566..409W} {566, 409}

\bibitem[\protect\citeauthoryear{{Winn} \& {Fabrycky}}{{Winn} \&
  {Fabrycky}}{2015}]{WinnFabrycky2015}
{Winn} J.~N.,  {Fabrycky} D.~C.,  2015, \mn@doi [\araa]
  {10.1146/annurev-astro-082214-122246}, \href
  {http://adsabs.harvard.edu/abs/2015ARA%26A..53..409W} {53, 409}

\bibitem[\protect\citeauthoryear{{de Boer} et~al.,}{{de Boer}
  et~al.}{2016}]{2016A&A...595A.114D}
{de Boer} J.,  et~al., 2016, \mn@doi [\aap] {10.1051/0004-6361/201629267},
  \href {https://ui.adsabs.harvard.edu/#abs/2016A&A...595A.114D} {595, A114}

\bibitem[\protect\citeauthoryear{{de Gregorio-Monsalvo} et~al.,}{{de
  Gregorio-Monsalvo} et~al.}{2013}]{deGregorioMonsalvo2013}
{de Gregorio-Monsalvo} I.,  et~al., 2013, \mn@doi [\aap]
  {10.1051/0004-6361/201321603}, \href
  {https://ui.adsabs.harvard.edu/#abs/2013A&A...557A.133D} {557, A133}

\bibitem[\protect\citeauthoryear{{van Boekel} et~al.,}{{van Boekel}
  et~al.}{2017}]{2017ApJ...837..132V}
{van Boekel} R.,  et~al., 2017, \mn@doi [\apj] {10.3847/1538-4357/aa5d68},
  \href {https://ui.adsabs.harvard.edu/#abs/2017ApJ...837..132V} {837, 132}

\bibitem[\protect\citeauthoryear{{van Dishoeck} \& {Black}}{{van Dishoeck} \&
  {Black}}{1988}]{1988ApJ...334..771V}
{van Dishoeck} E.~F.,  {Black} J.~H.,  1988, \mn@doi [\apj] {10.1086/166877},
  \href {http://adsabs.harvard.edu/abs/1988ApJ...334..771V} {334, 771}

\bibitem[\protect\citeauthoryear{{van Terwisga} et~al.,}{{van Terwisga}
  et~al.}{2018}]{2018A&A...616A..88V}
{van Terwisga} S.~E.,  et~al., 2018, \mn@doi [\aap]
  {10.1051/0004-6361/201832862}, \href
  {http://adsabs.harvard.edu/abs/2018A%26A...616A..88V} {616, A88}

\makeatother
\end{thebibliography}



\appendix

\section{An in-depth look at the mass evolution}

\label{sec:mass_evol_appendix}

\begin{figure}
\includegraphics[width=\columnwidth]{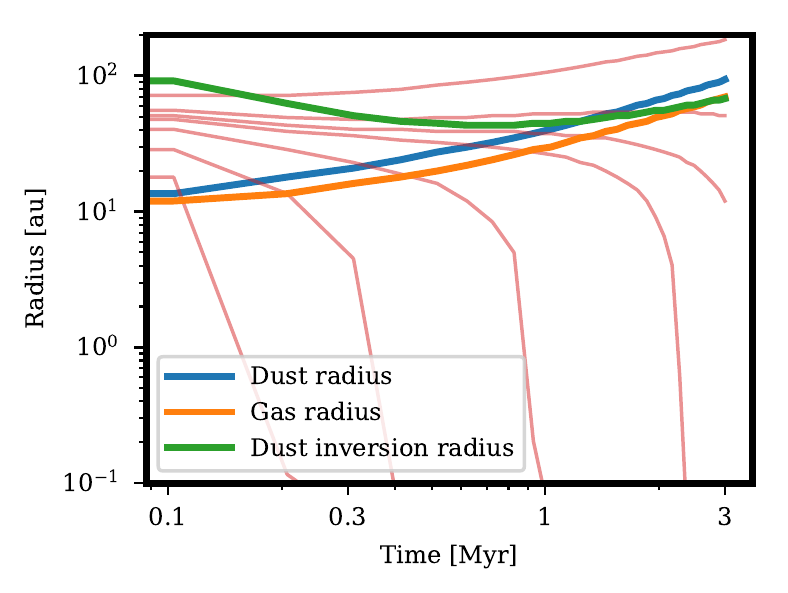}
\caption{Evolution of the gas, dust and inversion (see text) radius as a function of time for the fiducial model. The light red lines are the lagrangian trajectories of dust particles with different initial radii.}
\label{fig:rdisc_0.001_0.1_10}
\end{figure}

\begin{figure}
\includegraphics[width=\columnwidth]{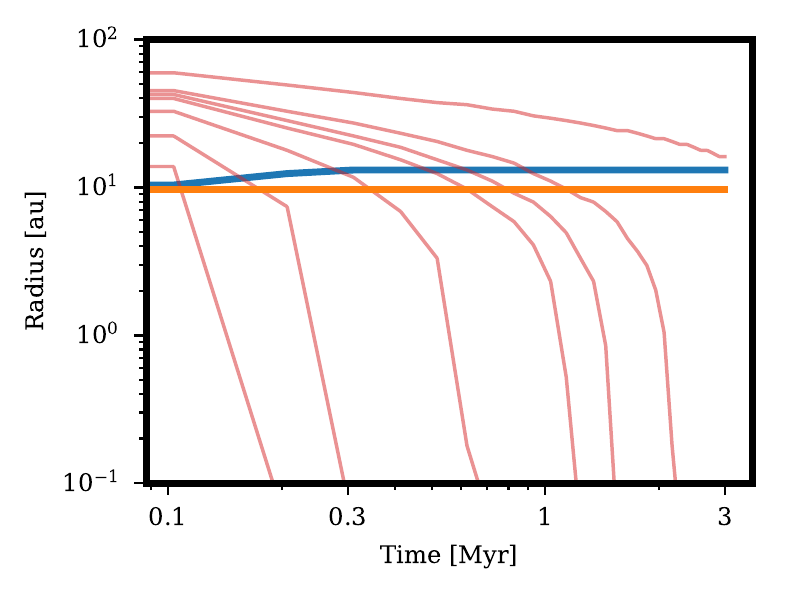}
\caption{Same as \autoref{fig:rdisc_0.001_0.1_10}, but where the gas disc is not viscously evolving. In this case the dust disc does not expand with time, showing that the viscosity of the gas is the reason why the dust disc expands. Note also how the dust trajectories are always directed inwards, yet the mass radius stays approximately constant.}
\label{fig:rdisc_no_visc_0.001_0.1_10}
\end{figure}

\begin{figure}
\centering
\includegraphics[width=0.45\textwidth]{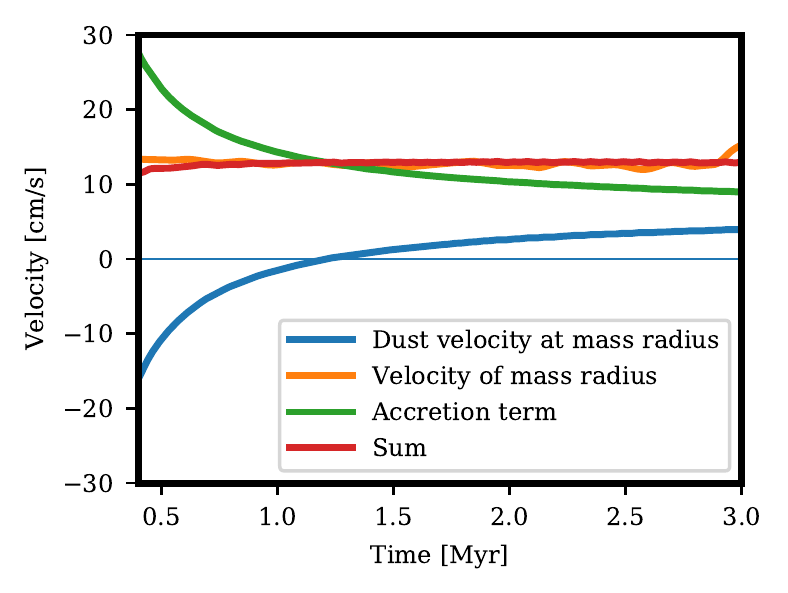}
\caption{Time evolution of the two terms in equation \eqref{eq:v_dissect}. The plot also shows the time derivative of the dust mass radius, compared with the sum of the two terms; as expected there is excellent agreement between the two.} 
\label{fig:v_analysis}
\end{figure}


We have already shown in \autoref{fig:dust_sigma_vel} the dust velocity at a given time, showing that there is one part of the disc moving outwards. To provide a more complete picture of how the dust velocity varies as a function of the time and space coordinates, in \autoref{fig:rdisc_0.001_0.1_10} we plot with the light red lines the lagrangian trajectories for different initial radii for the fiducial model. This clearly shows that particles starting at small radii eventually drift onto the star. Particles with larger initial radii instead either stay stationary or move outwards as a result of the outwards velocity in the outer part of the disc; these particles are the ones driving the expansion of the dust radius.

The behaviour of the dust is reminiscent of the behaviour of the gas disc, but with an important difference. Also in the gas component, as can be shown analytically studying the \citet{LyndenBellPringle74} solution, at any given time the inner part of the disc is moving inwards while the outer part is moving outwards. It is the expansion of the outer part that drives viscous spreading. However, in contrast with the gas, the dust at the mass radius moves \textit{inwards}. This is shown in \autoref{fig:rdisc_0.001_0.1_10} by the green line, which shows at any given time  the radius separating the inwards from outwards moving region (``\textit{dust inversion radius}'' for brevity). It can be seen how this radius initially moves inwards, which we interpret as due to an initial rapid phase of grain growth and drift, but eventually moves outwards as a significant amount of dust is accreted on the star and the grains become smaller. Only after $\sim$1.5 Myr the dust mass radius catches up with the dust inversion radius, i.e, the dust at the mass radius is instantaneously moving outwards. It follows that for a long part of the evolution most of the mass in the disc is moving towards the star, yet the disc overall expands.



For completeness, we show also the case without viscosity in \autoref{fig:rdisc_no_visc_0.001_0.1_10}. In this case we do not plot the dust inversion radius: the dust velocity is always directed inwards, since the viscous contribution is lacking. Despite the fact that at any given time all the mass in the disc is moving inwards, the mass radius is approximately constant.



The analysis of these plots clearly shows that the velocity of the mass radius differs significantly from the instantaneous velocity of the dust located at the mass radius. Indeed, it is instructive to derive a relation expressing how the mass radius $r_\mathrm{s}$ corresponding to a given fraction $f$ of the total mass evolves with time. It can be shown (see Appendix \ref{sec:half-evol}) that
\begin{equation}
\frac{\mathrm{d} r_\mathrm{s}}{\mathrm{d} t} =  \frac{- \dot{M} (r_\mathrm{s}) + (1-f) \dot{M}(r_\ast)  }{2 \pi r_\mathrm{s} \Sigma(r_\mathrm{s})},
\label{eq:v_dissect}
\end{equation}
where 
\begin{equation}
\dot{M} (r)=2\pi r v_r \Sigma
\label{eq:mdot}
\end{equation}
is the mass accretion rate at any given radius and we have denoted with $r_\ast$ the radius of the star (or for our purposes, the inner boundary of our grid). The expression shows that the mass radius evolves due to two contributions. The first term (``velocity term'') is the instantaneous velocity at the mass radius. The additional term (``accretion term'') is present because the disc is losing mass at the inner boundary, causing an outward shift in the mass radius. The competition between this outward term and the inwards dust velocity determines whether the disc expands or shrinks.

\autoref{fig:v_analysis} shows the evolution of the two terms in equation \eqref{eq:v_dissect} in the left panel. As previously noted, the dust at the mass radius is moving inwards for roughly half of the simulation. The reason why the mass radius increases at all times is the accretion term, which always dominates the evolution even when the velocity term becomes positive. This shows that radial drift is a victim of its own success: by promoting a fast radial drift it also causes a large accretion term. In other words, drift rapidly removes from the disc the grains that are drifting the fastest, so that the grains in the outer disc are the ones dominating the radius evolution. Finally, the plot also shows the time derivative of the mass radius computed from the solution in comparison to the two terms, showing that the two agree.


\section{Derivation of mass radius evolution}
\label{sec:half-evol}

In this section we derive \autoref{eq:v_dissect}, which expresses the time-derivative of the mass radius. The formal definition of the mass radius $r_s$ is given implicitly by the relation
\begin{equation}
\int_{r_\ast}^{r_s (t)} 2 \pi r \Sigma(r,t) \mathrm{d}r=f \int_{r_\ast}^{\infty} 2 \pi r \Sigma(r,t) \mathrm{d}r = f M,
\end{equation}
where $M$ is the total disc mass and $0<f<1$ the chosen fraction of the total disc mass. By taking the derivative with respect to time of this expression, we obtain
\begin{equation}
\frac{\partial}{\partial t} \int_{r_\ast}^{r_s (t)} 2 \pi r \Sigma(r,t) \mathrm{d}r = 2 \pi r_s \Sigma(r_s,t) \frac{\partial r_s}{\partial t} + \int_{r_\ast}^{r_s (t)} 2 \pi r \frac{\partial \Sigma(r,t)}{\partial t} \mathrm{d}r.
\label{eq:rs_dt}
\end{equation}
To simplify this expression, we note that the time-derivative of the disc surface density is given by the mass continuity equation
\begin{equation}
\frac{\partial \Sigma}{\partial t} + \frac{1}{r} \frac{\partial}{\partial r} \left( r \Sigma v_r \right) = 0.
\end{equation}
Substituting into \autoref{eq:rs_dt} and using the definition of mass accretion rate given by \autoref{eq:mdot} yields
\begin{equation}
f \frac{\partial M}{\partial t} = 2 \pi r_s \Sigma(r_s,t) \frac{\partial r_s}{\partial t} + (\dot{M} (r_s) - \dot{M} (r_\ast) ).
\end{equation}
We can rewrite the left-hand side using yet again the continuity equation, obtaining
\begin{equation}
-f \dot{M} (r_\ast) = 2 \pi r_s \Sigma(r_s,t) \frac{\partial r_s}{\partial t} + (\dot{M} (r_s) - \dot{M} (r_\ast) ),
\end{equation}
which can be re-arranged to give \autoref{eq:v_dissect}.

\section{Comparison between flux and mass evolution}
\begin{figure*}

\parbox{\columnwidth}{
\includegraphics[width=\columnwidth]{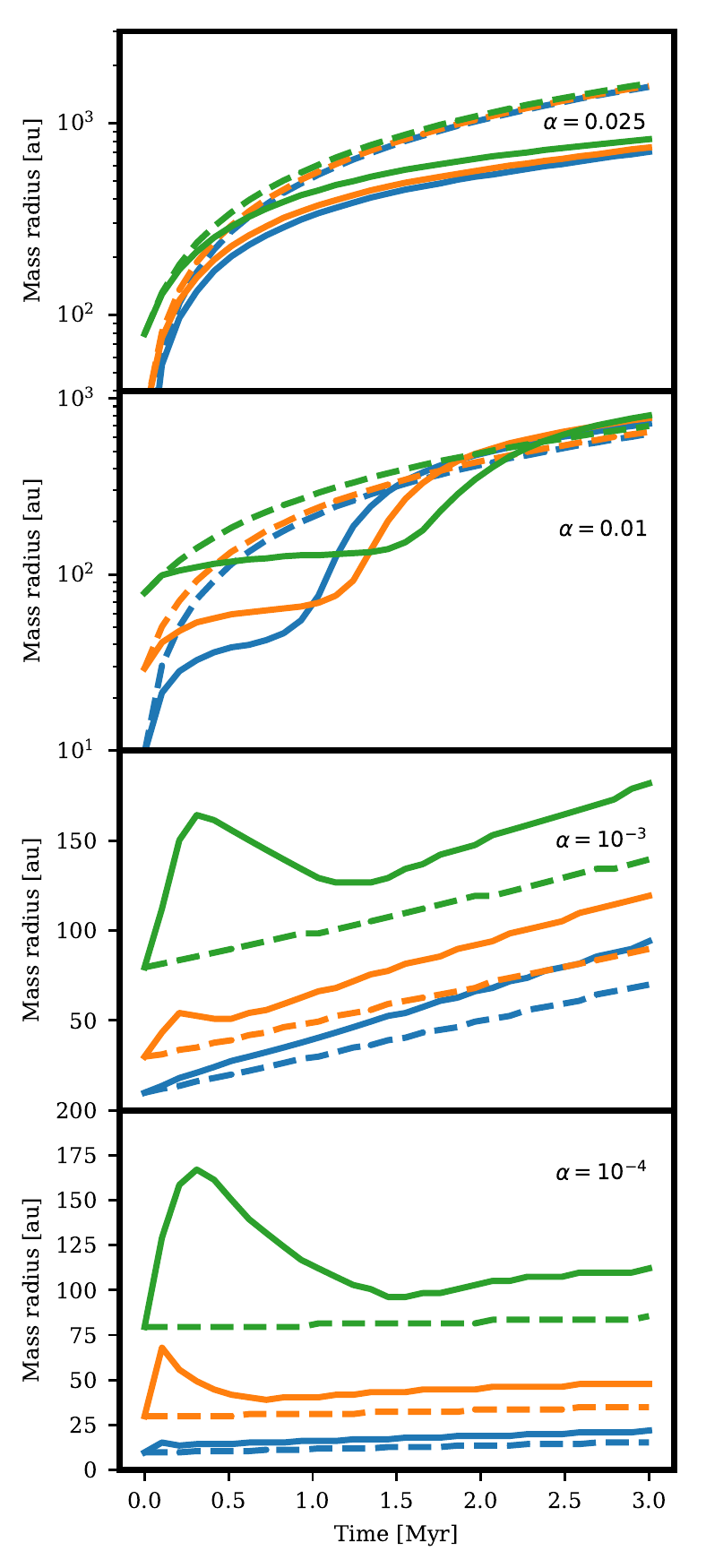}
\centering
\includegraphics[]{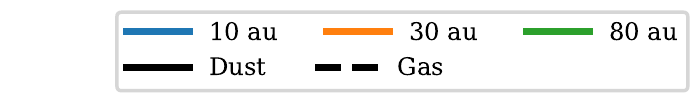}
}
\parbox{\columnwidth}{
\includegraphics[width=\columnwidth]{{r_t_flux_panel}.pdf}
\centering
\includegraphics[]{legend_figure13.pdf}
}

\caption{We plot side by side the evolution of the flux and mass radii to allow for an easier comparison. \textbf{Left panel}: evolution of the mass radius with time for different viscosities and initial radii. \textbf{Right panel}: Evolution of the flux radii with time.}
\label{fig:r_t_mass_flux_comparison}
\end{figure*}

\bsp	
\label{lastpage}
\end{document}